\font\big=cmbx10 scaled \magstep4

\newcommand\hmpc{h^{-1}{\rm Mpc}}
\newcommand\hkpc{h^{-1}{\rm kpc}}
\newcommand\khmpc{h{\rm Mpc}^{-1}}
\newcommand\hmsun{h^{-1}M_{\sun}}
\newcommand\whsp{\phantom{222}}
\newcommand\whspb{\phantom{22222}}
\newcommand\LCDMa{$\Lambda$CDM1}
\newcommand\SCDMa{SCDM1}
\newcommand\OCDMa{OCDM1}
\newcommand\TCDMa{$\tau$CDM1a}
\newcommand\TCDMab{$\tau$CDM1b}
\newcommand\LCDMb{$\Lambda$CDM2}
\newcommand\SCDMb{SCDM2}
\newcommand\TCDMb{$\tau$CDM2}
\newcommand\OCDMb{OCDM2}
\newcommand\LCDM{$\Lambda{\rm CDM}$}
\newcommand\tCDM{$\tau{\rm CDM}$}
\newcommand\SCDM{SCDM}
\newcommand\OCDM{OCDM}
\newcommand\kms{{\rm kms}^{-1}}
\newcommand\Mpc{{\rm Mpc}}
\newcommand\vrad{$v_{21}$}
\newcommand\vradisp{$v_{\parallel}^2$}
\newcommand\vtandisp{$v_{\perp}^2$}
\newcommand\vradispo{$v_{\parallel}$}
\newcommand\vtandispo{$v_{\perp}$}

\newcommand\ddd{{\rm d}}
\newcommand\ts{\scriptscriptstyle\rm}
\newcommand\etal{{\it et al.}\ \rm}
\documentstyle[11pt,aaspp4]{article}
\lefthead{Jenkins \etal.}
\righthead{Evolution of structure.}

\begin{document}

\title{Evolution of structure in cold dark matter \\
universes.}

\author{A. Jenkins\altaffilmark{1}, C. S. Frenk\altaffilmark{1},
F. R. Pearce\altaffilmark{1}\altaffilmark{2}, 
P. A. Thomas\altaffilmark{2}, J. M. Colberg\altaffilmark{3},
S. D. M. White\altaffilmark{3}, $\phantom{xxxxx}$ \ H. M. P. Couchman\altaffilmark{4},
J. A. Peacock\altaffilmark{5}, G. Efstathiou\altaffilmark{6}\altaffilmark{7}
and A. H. Nelson\altaffilmark{8} \ \ \ \ \ \ $\phantom{xxxxxxxxxxx}$ 
(The Virgo Consortium)}

\affil{}

\altaffiltext{1}{Dept Physics, South Road, University of Durham, DH1 3LE} 
\altaffiltext{2}{CPES, University of Sussex, Falmer, Brighton BN1 9QH}
\altaffiltext{3}{Max-Planck Inst. for Astrophysics, Garching, Munich,
D-85740, Germany}
\altaffiltext{4}{Dept of Astronomy, University of Western Ontario, London,
Ontario N6A 3K7, Canada}
\altaffiltext{5}{Royal observatory, Blackford Hill, Edinburgh, EH9 3HJ}
\altaffiltext{6}{Dept Physics, Nuclear Physics Building, Keble Road,
Oxford, OX1 3RH}
\altaffiltext{7}{Institute of Astronomy, Madingley Road, Cambridge, CB3 OHA}
\altaffiltext{8}{Dept of Physics and Astronomy, University of Wales, PO Box
913, Cardiff CF2 3YB}

\clearpage

\setcounter{footnote}{1}
\begin{abstract}

We present an analysis of the clustering evolution of dark matter in four
cold dark matter (CDM) cosmologies. We use a suite of high resolution,
17-million particle, N-body simulations which sample volumes large enough
to give clustering statistics with unprecedented accuracy. We investigate a
flat model with $\Omega_0=0.3$, an open model also with $\Omega_0=0.3$, and
two models with $\Omega=1$, one with the standard CDM power spectrum and
the other with the same power spectrum as the $\Omega_0=0.3$ models. In all
cases, the amplitude of primordial fluctuations is set so that the models
reproduce the observed abundance of rich galaxy clusters by the present
day. We compute mass two-point correlation functions and power spectra over
three orders of magnitude in spatial scale and find that in all our
simulations they differ significantly from those of the observed galaxy
distribution, in both shape and amplitude. Thus, for any of these models to
provide an acceptable representation of reality, the distribution of
galaxies must be biased relative to the mass in a non-trivial,
scale-dependent, fashion. In the $\Omega=1$ models the required bias is
always greater than unity, but in the $\Omega_0=0.3$ models an ``antibias"
is required on scales smaller than $\sim 5 \hmpc$. The mass correlation
functions in the simulations are well fit by recently published analytic
models. The velocity fields are remarkably similar in all the models,
whether they be characterised as bulk flows, single-particle or pairwise
velocity dispersions. This similarity is a direct consequence of our
adopted normalisation and runs contrary to the common belief that the
amplitude of the observed galaxy velocity fields can be used to constrain
the value of $\Omega_0$. The small-scale pairwise velocity dispersion of
the dark matter is somewhat larger than recent determinations from galaxy
redshift surveys, but the bulk flows predicted by our models are broadly in
agreement with most available data.
\end{abstract}

\clearpage
\section{Introduction}

Cosmological N-body simulations play a pivotal role in the study of 
the formation of cosmic structure. In this methodology, initial
conditions are set at some early epoch by using linear theory to calculate
the statistical properties of the fluctuations. Such a calculation
requires some specific mechanism for generating primordial structure,
together with assumptions about the global cosmological parameters and the
nature of the dominant dark matter component. N-body simulations are then
used to follow the later evolution of the dark matter into the nonlinear
regime where it can be compared with the large-scale structure in galaxy
surveys. This general picture was developed fully in the early 1980s,
building upon then novel concepts like the inflationary model of the early
universe and the proposition that the dark matter is non-baryonic. In the
broadest sense, it was confirmed in the early 1990s with the discovery of
fluctuations in the temperature of the microwave background radiation
(Smoot \etal 1992). The plausibility of the hypothesis that the dark
matter is non-baryonic has strengthened in recent years, as the gap
between the upper limit on the density of baryons from Big Bang
nucleosynthesis considerations (e.g. \cite{Tytler96}) and the lower limit
on the total mass density from dynamical studies (e.g. \cite{Carlberg97})
has become more firmly established.

Cosmological N-body simulations were first employed to study the
large-scale evolution of dark matter on mildly nonlinear scales, a regime
which can be accurately calculated using relatively few particles.
Highlights of these early simulations include the demonstration of the
general principles of nonlinear gravitational clustering (\cite{Gat});
evidence that scale-free initial conditions evolve in a self-similar way
(\cite{EE81}; \cite{EDFW}), while truncated power 
spectra develop large-scale pancakes and filaments (\cite{KS83};
\cite{cm83}; \cite{FWD83}); and the rejection of the proposal 
that the dark matter consists of light massive neutrinos (\cite{White83a}; 
\cite{White84}).

During the mid-1980s, N-body simulations were extensively used to 
explore the hypothesis, first elaborated by Peebles (1982), that the
dark matter consists of cold collisionless particles. This hypothesis --
the cold dark matter (CDM) cosmology -- has survived the test of time and
remains the basic framework for most contemporary cosmological work. The
clustering evolution of dark matter in a CDM universe was first studied in
detail using relatively small N-body simulations (\cite{DEFW}, hereafter
DEFW; Frenk \etal 1985, 1988, 1990; White \etal 1987a, 1987b; \cite{Fm85}).
In particular, DEFW concluded, on the basis of 32768-particle simulations,
that the simplest (or standard) version of the theory in which the mean
cosmological density parameter $\Omega=1$, and the galaxies share the same
statistical distribution as the dark matter, was inconsistent with the low
estimates of the rms pairwise peculiar velocities of galaxies which had
been obtained at the time from the CfA redshift survey (\cite{MDP83}). They
showed that much better agreement with the clustering data available at the
time could be obtained in an $\Omega=1$ CDM model if the galaxies were
assumed to be biased tracers of the mass, as in the ``high peak model" of
galaxy formation (\cite{Kaiser84}; \cite{BBKS}). They found that an equally
successful CDM model could be obtained if galaxies traced the mass but
$\Omega_0 \simeq 0.2$, and the geometry was either open or flat. Many of
the results of this first generation of N-body simulations have been
reviewed by Frenk (1991).

Following the general acceptance of cosmological simulations as a useful
technique, the subject expanded very rapidly. To mention but a few examples
in the general area of gravitational clustering, further simulations have
re-examined the statistics of the large-scale distribution of cold dark
matter (e.g. \cite{Park91}; \cite{Gelb94a}, 1994b; \cite{Klypin96};
\cite{Cole97};\cite{Zurek94}), confirming on the whole, the results of the 
earlier, smaller calculations. Large simulations have been used to
construct ``mock" versions of real galaxy surveys
(e.g. \cite{White87b};
\cite{Park94}; \cite{Moore94}), or to carry out ``controlled experiments''
designed to investigate specific effects such as non-gaussian initial
conditions (\cite{WC92}) or features in the power spectrum (\cite{MS93}).
Some attempts have been made to address directly the issue of where
galaxies form by modelling the evolution of cooling gas gravitationally
coupled to the dark matter (e.g. \cite{Carlberg90}; \cite{Cen92},
\cite{KHW92}; \cite{ESD94}; \cite{Jenk97}). The success of the N-body
approach has stimulated the development of analytic approximations to
describe the weakly nonlinear behavior, using, for example, second order
perturbation theory (e.g. \cite{Bern94};
\cite{Bouchet95}), as well as Lagrangian approximations to the fully
nonlinear regime (\cite{HKLM91}; \cite{JMW}; \cite{BG96}; 
\cite{PD94}, 1996; \cite{Pad96}).

Steady progress has also been achieved on the observational front with the
completion of ever larger galaxy surveys. The first real indication that
the galaxy distribution on large scales differs from that predicted by the
standard cold dark matter model was furnished by the APM survey which
provided projected positions and magnitudes for over a million galaxies.
The angular correlation function of this survey has an amplitude that
exceeds the theoretical predictions by a factor of about 3 on scales of 20
to $30\hmpc$ (\cite{MAD90}). This result has been repeatedly confirmed in
redshift surveys of IRAS (e.g. \cite{Efst90}; \cite{Saunders90};
\cite{Tadros95}), and optical galaxies (e.g. \cite{Vog92}; 
\cite{Tadros96}; \cite{Tucker97}; \cite{Ratcliffe97}.) Modern redshift 
surveys have also allowed better estimates of the peculiar velocity field
of galaxies in the local universe. The original measurement of the pairwise
velocity dispersion (which helped motivate the concept of biased galaxy
formation in the first place) has been revised upwards by Mo, Jing and
B\"orner (1993) and Sommerville, Davis \& Primack (1997), but Marzke \etal
(1995) and Mo, Jing \& B\"orner (1996) have argued that such pairwise
statistics are not robust when determined from relatively small redshift
surveys. The Las Campanas redshift survey is, perhaps, the first which is
large enough to give a robust estimate of these statistics
(\cite{JMB97}). Surveys of galaxy distances are also now beginning to map
the local mean flow field of galaxies out to large distances (e.g.
\cite{LBSS88}; \cite{Courteau93}; \cite{Mould93}; Dekel \etal\ 1997; 
Giovanelli 1997; \cite{Saglia97}; \cite{Willick97}.) 
Both pairwise velocity dispersions and mean flows allow an estimate of the
parameter combination $\beta\equiv\Omega_0^{0.6}/b$ (where $b$ is the
biasing parameter defined in Section~\ref{correlation-section}); recent
analyses seem to be converging on values of $\beta$ around 0.5.

In this paper we present results from a suite of very large,
high-resolution N-body simulations. Our primary aim is to extend the N-body
work of the 1980s and early 1990s by increasing the dynamic range of the
simulations and calculating the low-order clustering statistics of the dark
matter distribution to much higher accuracy than is possible with
smaller calculations. Our simulations follow nearly $17$ million particles,
with a spatial resolution of a few tens of kiloparsecs and thus probe 
the strong clustering regime whilst correctly including large-scale
effects. Such improved theoretical predictions are a necessary counterpart
to the high precision attainable with the largest galaxy datasets like
the APM survey and particularly the forthcoming generation of redshift
surveys, the Sloan (\cite{GW95}) and 2-degree field
(http:$\backslash\backslash$
www.ast.cam.ac.uk$\backslash$~2dFgg$\backslash$) projects. Our simulations
do not address the issue of where galaxies form. They do, however, reveal
in quantitative detail the kind of biases that must be imprinted during the
galaxy formation process if any of the models is to provide an acceptable
match to the galaxy clustering data. We examine four versions of the cold
dark matter theory including, for the first time, the \tCDM\ model.  This
has $\Omega=1$ but more power on large scales than the standard version and
offers an attractive alternative to the standard model if $\Omega=1$.  We
focus on high precision determinations of the spatial and velocity
distributions and also carry out a comparison of the simulation results
with the predictions of analytic clustering models.

Many of the issues we discuss in this paper have been addressed previously
using large N-body simulations. Our study complements and supersedes
aspects of this earlier work because our simulations are significantly
larger and generally have better resolution than earlier simulations and
also because we investigate four competing cosmological models in a uniform
manner. Thus, for example, Gelb and Bertschinger (1994b) studied the
standard $\Omega=1$ CDM model but most of their simulations had
significantly poorer spatial resolution than ours and the one with similar
resolution had only 1\% of the volume.  Klypin \etal (1996) simulated a
low-$\Omega_0$ flat CDM model with a mass resolution at least 10 times
poorer than ours or in volumes that were too small to properly include the
effects of rare objects. These simulations missed a number of subtle,
but nevertheless important, effects that are revealed by our larger 
simulations. 
Our analysis has some features in common with the recent work of Cole \etal
(1997) who simulated a large suite of cosmologies in volumes that are
typically three times larger than ours, but have 3-6 times fewer particles
and an effective mass resolution an order of magnitude less than
ours. Their force resolution is also a factor of three times worse that
ours. While Cole \etal focussed on models in which the primordial
fluctuation amplitude is normalised using the inferred amplitude of the
COBE microwave background fluctuations, our models are normalized so that
they all give the observed abundance of rich galaxy clusters by the present
day. Our choice of normalisation is motivated and explained in
Section~\ref{sim-section}. 

This study is part of the programme of the ``Virgo consortium," an
international collaboration recently constituted with the aim of carrying
out large N-body and N-body/gasdynamic simulations of large-scale
structure and galaxy formation, using parallel supercomputers in Germany
and the UK. Some of our preliminary results are discussed in Jenkins \etal
(1997) and further analysis of the present simulations may be found in
Thomas \etal (1997).

The cosmological parameters of our models are described in
Section~\ref{cos_mods}\ and their numerical details in
Section~\ref{sim-section}. Colour images illustrating the evolution of
clustering in our simulations are presented in
Section~\ref{slices-section}. The evolution of the mass correlation
functions and power spectra are discussed, and compared with observations,
in Sections~\ref{correlation-section} and~\ref{powersp-section}. We 
compare these clustering statistics with analytic models for the nonlinear
evolution of correlation functions and power spectra in
Section~\ref{pred-section}. The present day velocity fields, both bulk
flows and pairwise dispersions, are discussed in
Section~\ref{velocityfield}.  Our paper concludes in Section~9 with a
discussion and summary (including a table) of our main results.

\section{Cosmological models\label{cos_mods}}

We have simulated evolution in four CDM cosmologies with parameters
suggested by a variety of recent observations. The shape of the CDM power
spectrum is determined by the parameter, $\Gamma$,
(c.f. equation~\ref{powspec} below); observations of galaxy clustering,
interpreted via the assumption that galaxies trace the mass, indicate a
value $\Gamma\simeq 0.2$ (Maddox \etal 1990, 1996; \cite{Vog92}). In the
standard version of the theory, $\Gamma=\Omega_0 h$,\footnote[1]{Here and
below we denote Hubble's constant $H_0$ by $h=H_0/100\,\kms\Mpc^{-1}$} which 
corresponds, for low baryon density, to the standard assumption that only
photons and three massless species of neutrinos and their antiparticles
contribute to the relativistic energy density of the Universe at late
times. For a given $\Omega$ and $h$, smaller values of $\Gamma$ are
possible, but this requires additional physics, such as late decay of the
(massive) $\tau$-neutrino to produce an additional suprathermal background of
relativistic e- and $\mu$-neutrinos at the present day (\cite{WGS95}).  This
has the effect of delaying the onset of matter domination, leading to a
decrease in the effective value of $\Gamma$.

In addition to observations of large-scale structure, a second
consideration that has guided our choice of cosmological models is the
growing evidence in favour of a value of $\Omega_0$ around 0.3. The
strongest argument for this is the comparison of the baryon fraction in
rich clusters with the universal value required by Big Bang nucleosynthesis
(\cite{White93}; \cite{WF95}; \cite{Evrard97}).  The recently determined
abundance of hot X-ray emitting clusters at $z\simeq 0.3$ also indicates a
similar value of $\Omega_0$ (\cite{Henry97}.) The strength of these tests
lies in the fact that they do not depend on uncertain assumptions regarding
galaxy formation. Nevertheless, they remain controversial and so, in
addition to cosmologies with $\Omega_0=0.3$, we have also simulated models
with $\Omega=1$.

Three of our simulations have a power spectrum shape parameter,
$\Gamma=0.21$. One of these (\LCDM) has $\Omega_0=0.3$ and the flat
geometry required by standard models of inflation,
i.e. $\lambda\equiv\Lambda/(3H^2)=0.7$ (where $\Lambda$ is the cosmological
constant and $H$ is Hubble's constant). The second model (\OCDM) also has
$\Omega_0=0.3$, but $\Lambda=0$. In both these models we take $h=0.7$, 
consistent with a number of recent determinations 
(\cite{Kenn95}). Our third model with $\Gamma=0.21$ (\tCDM) has $\Omega=1$
and $h=0.5$; this could correspond to the decaying neutrino model mentioned
above.  Finally, our fourth model is standard CDM (\SCDM) which has
$\Omega=1$, $h=0.5$, and $\Gamma=0.5$. Thus, two of our models (\LCDM\ and
\OCDM) differ only in the value of the cosmological constant; two others
(\LCDM\ and \tCDM) have the same power spectrum and geometry but different
values of $\Omega_0$; and two more (\tCDM\ and \SCDM) differ only in the
shape of the power spectrum.

Having chosen the cosmological parameters, we must now set the amplitude of
the initial fluctuation spectrum. DEFW did this by requiring that the
slope of the present day two-point galaxy correlation function in the
simulations should match observations. This was a rather crude method, but
one of the few practical alternatives with the data available at the
time. The discovery of fluctuations in the temperature of the microwave
background radiation by COBE offered the possibility of normalising the
mass fluctuations directly by relating these to the measured temperature
fluctuations on large scales. In practice, however, the large
extrapolation required to predict the amplitude of fluctuations on scales
relevant to galaxy clustering from the COBE data makes this procedure
unreliable because it depends sensitively on an uncertain assumption about
the slope of the primordial power spectrum. A further source of uncertainty is
the unknown contribution to the COBE signal from tensor (rather than
scalar) modes. In spite of these uncertainties, it is remarkable that the
normalisation inferred from the simplest possible interpretation of the
COBE data is within about a factor of 2 of the normalisation inferred for
standard CDM by DEFW from galaxy clustering considerations.

A more satisfactory procedure for fixing the amplitude of the initial mass
fluctuations is to require that the models should match the observed
abundance of galaxy clusters. The distribution of cluster abundance,
characterised by mass, X-ray temperature or some other property, declines
exponentially and so is very sensitive to the normalisation of the power
spectrum (\cite{FWED90}). Using the observed cluster abundance to
normalise the power spectrum has several advantages.  Firstly, it is based
on data which are well matched to the scales of interest; secondly, it
gives the value of $\sigma_8$ (the linearly extrapolated rms of the density
field in spheres of radius $8\hmpc$) with only a weak dependence on the
shape of the power spectrum if $\Omega<1$ and no dependence at all if
$\Omega=1$ (\cite{WEF93}); thirdly, it does not require a particularly
accurate estimate of the abundance of clusters because of the strong
sensitivity of abundance on $\sigma_8$. The disadvantage of this method is
that it is sensitive to systematic biases arising from inaccurate
determinations of the particular property used to
characterize the abundance.  However, the consistency of the estimates of
$\sigma_8$ when the abundance of clusters is characterized by total mass
(\cite{HA91}), by mass within the Abell radius (\cite{WEF93}), or by the
X-ray temperature of the intracluster medium (\cite{Eke96};
\cite{Viana96}) suggests that systematic effects are likely to be
small.

We adopt the values of $\sigma_8$ recommended by Eke, Cole \& Frenk (1996)
from their analysis of the local cluster X-ray temperature function. This 
requires: 
\begin{equation}
\sigma_8 = 
(0.52\pm0.04)\Omega_0^{-0.52+0.13\Omega_0}\ \ \ ({\rm flat\ \ models})
\label{norm1}
\end{equation}
or 
\begin{equation}
 \sigma_8 =
(0.52\pm0.04)\Omega_0^{-0.46+0.1\Omega_0}\ \ \ ({\rm open\ \ models}) 
\label{norm2}
\end{equation}
These values of $\sigma_8$ are consistent with those obtained from the slightly
different analyses carried out by White, Efstathiou \& Frenk (1993), Viana
\& Liddle (1996) and Henry (1997). 

The resulting values of $\sigma_8$ for our simulations are listed in
Table~1. For reference, these values may be compared to those required by
the COBE data under the simplest set of assumptions, namely that the
primordial power spectrum is a power-law with exponent $n=1$ (the
Harrison-Zel'dovich spectrum) and that there is no contribution at all from
tensor modes. For our chosen cosmologies, the 4-year COBE-DMR data imply
values of $\sigma_8$ of 1.21, 0.45, 1.07, 0.52 (\cite{Gorski95},
\cite{Ratra97}) for \SCDM, \tCDM, \LCDM, and \OCDM\ respectively. Thus, 
our \tCDM\ and \LCDM\ models are roughly consistent with the conventional
COBE normalisation, but our adopted normalisations for the
\SCDM\ and \OCDM\ models are $\sim 40\%$ lower and $\sim 60\%$ higher 
respectively than the COBE values. These numbers are consistent with those
obtained by Cole \etal (1997) from their grid of large COBE-normalised
cosmological N-body simulations with different parameter values. As may be seen
from their Figure~4, there is only a small region of parameter space in
which the conventional COBE-normalised CDM models produce the correct
abundance of clusters. Flat models require $0.25\le\Omega_0\le0.4$ while
open models require $0.4\le\Omega_0\le0.5$.

To summarize, we have chosen to simulate four cosmological models which are
of interest for a variety of reasons. Our three flat models are consistent
with standard inflationary theory and our open model can be motivated by
the more exotic ``open bubble'' version of this theory (\cite{GB97}). By
construction, all our models approximately reproduce the observed abundance
of rich galaxy clusters. The \LCDM\ model has a value of $\Omega_0$ in line
with recent observational trends and a value of $\Gamma$ that is close to
that inferred from galaxy clustering. It has the additional advantages that
its normalisation agrees approximately with the conventional COBE
normalisation and, for our adopted value of $H_0$, it has an age that is
comfortably in accord with traditional estimates of the ages of globular
clusters (\cite{Renzini96}, but see \cite{Jimenez96}). The
\OCDM\ model shares some of these attractive features but allows us also to
investigate the effects of the cosmological constant on the dynamics of
gravitational clustering. Its normalisation is higher than required to
match the conventional COBE value, but this could be rectified by a modest
increase in $\Omega_0$ to about 0.4-0.5. The \tCDM\ model is as well
motivated by galaxy clustering data as are the low-$\Omega_0$ models and
has the advantage that it allows us to investigate the dynamical effects of
changing $\Omega_0$ while keeping the shape of the initial power spectrum
fixed. Finally, the traditional
\SCDM\ model is an instructive counterpart to its \tCDM\ variant.
 
\section{The Simulations\label{sim-section}}

Our simulations were carried out using a parallel, adaptive
particle-particle/particle-mesh code developed by the Virgo consortium
(\cite{Pearce95}, \cite{PC97}). This is identical in operation to the
publicly released serial version of ``Hydra'' (\cite{CPT96}; see
\cite{CTP95} for a detailed description.)  The simulations presented in
this paper are the first carried out by the Virgo consortium and were
executed on either 128 or 256 processors of the Cray T3Ds at the Edinburgh
Parallel Computing Centre and the Rechenzentrum, Garching.

The force calculation proceeds through several stages. Long range
gravitational forces are computed in parallel by smoothing the mass
distribution onto a mesh, typically containing $512^3$ cells, which is then
fast Fourier transformed and convolved with the appropriate Green's
function. After an inverse FFT, the forces are interpolated from the mesh
back to the particle positions. In weakly clustered regions, short range
(particle-particle) forces are also computed in parallel using the entire
processor set. Hydra recursively places additional higher resolution
meshes, or {\it refinements}, around clustered regions. Large refinements
containing over $\simeq10^5$ particles are executed in parallel by all
processors while smaller refinements, which fit within the memory of a
single processor, are most efficiently executed using a task farm approach.
The parallel version of Hydra employed in this paper is implemented in
CRAFT, a directive based parallel Fortran compiler developed for the Cray
T3D supercomputer (\cite{CRAFT}). We have checked that the introduction of
mesh refinements in high density regions does not introduce inaccuracies in
the computation by redoing our standard \tCDM\ simulation using a parallel
${\rm P}^3{\rm M}$ code (without refinements).  The two-point correlation
functions in these two simulations differed by less than 0.5\% over the
range $0.1\hmpc$ -- $5\hmpc$. 

%Global array addressing allows the programmer to treat
%a distributed memory machine like the Cray T3D as a shared memory machine.
%This, combined with a fast locking facility to prevent race conditions
%when two or more processors try to increment the same memory location,
%greatly reduces the difficulties involved in parallelising the method.
%However, a considerable amount of work was required to make the code run
%efficiently. Load imbalance amongst the processors as the particle
%distribution becomes increasingly clustered is the main problem. The
%performance of the code is described in xxx.

\placetable{tbl-1}

\subsection{Simulation details\label{rundetails}}

Initial conditions were laid down by imposing perturbations on an
initially uniform state represented by a ``glass" distribution of particles
generated by the method of White (1996). Using the algorithm
described by Efstathiou \etal (1985), based on the Zel'dovich
(1970) approximation, a Gaussian random field is set up by perturbing the
positions of the particles and assigning them velocities according to
growing mode linear theory solutions. Individual modes are assigned random
phases and the power for each mode is selected at random from an
exponential distribution with mean power corresponding to the desired
power spectrum $\Delta^2(k)$. 

Following Peebles' (1980) convention we define the dimensionless power
spectrum, $\Delta^2(k)$, as the power per logarithmic interval in
spatial frequency, $k$:
\begin{equation}
\Delta^2(k)\equiv {V\over (2\pi)^3}
\, 4\pi \,k^3\, |\delta_k|^2,
\end{equation}
where $|\delta_k|^2$ is the power density and $V$ is the volume. If the
primordial power spectrum is of the form $|\delta_k|^2\propto k^n$, then
the linear power spectrum at a later epoch is given by $\Delta^2(k)=k^{n+3}
T^2(k,t)$, where $T(k,t)$ is the transfer function. The standard inflationary
model of the early universe predicts that $n\simeq 1$ (Guth \& Pi 1982) and
we shall take $n=1$. For a cold dark matter model, the transfer function
depends on the values of $h$ and the mean baryon density $\Omega_b$. 
We use the approximation to the linear CDM power spectrum
given by Bond \& Efstathiou (1984), 
\begin{equation}\label{powspec}
 \Delta^2(k) = {Ak^4\over\bigg[1+\big[aq + (bq)^{3/2} 
+ (cq)^2\big]^{\nu}\bigg]^{2/\nu}}, 
\end{equation}
where $q = k/\Gamma$, $a = 6.4\hmpc$, $b = 3\hmpc$, $c = 1.7\hmpc$ and
$\nu=1.13$. The normalisation constant, $A$, is chosen by fixing the value
of $\sigma_8$ as discussed in Section~\ref{cos_mods}.

For our models, the analytic approximation of equation~(\ref{powspec})
provides a good approximation to the accurate numerical power spectrum
calculated by Seljak \& Zaldarriaga (1996) using their publicly
available code CMBFAST (http://arcturus.mit.edu:80/ $\sim$matiasz/
CMBFAST /cmbfast.html). For example, setting $h=0.7$ and
$\Omega_b=0.026$ in our \LCDM\ and \OCDM\ and normalizing to the same
value of $\sigma_8$, we find that the maximum difference at small
scales between the fit of equation~(\ref{powspec}) and the output of
CMBFAST is 13\% in power or 6\% in amplitude. These numbers are
smaller for a lower value of $\Omega_b$ or a small increase in
$h$. These differences are comparable to those induced by plausible
changes in $\Omega_b$ or $h$. (For example, for a \LCDM\ model, the
ratio of the $\sigma_8$-normalized CMBFAST power spectra for
$\Omega_b=0.01$\ and $\Omega_b=0.03$ respectively is 1.08 at the
Nyquist frequency of our simulation volumes ($k=3.36\khmpc$) and 0.85
at the fundamental frequency ($k=0.0262\khmpc$); if $\Omega_b$ is kept
fixed but $h$ is allowed to vary between 0.67 and 0.73, these ratios
become 1.08 and 0.9 respectively.)  Similarly, we set up our $\tau$CDM
model simply by changing the value of $\Gamma$ in
equation~(\ref{powspec}). This gives a satifactory fit provided that
the length-scale introduced in the power spectrum by the decay of the
$\tau$-neutrino is smaller than Nyquist frequency of the simulation
volume. This requires the mass of the decaying particle to be in
excess of about 10keV (\cite{BE91}).  Thus, over the range of
wavenumbers relevant to our simulations, equation~(\ref{powspec})
gives a good, but not perfect approximation to the true $\tau$CDM
power spectrum for a broad one-dimensional subset of the
two-dimensional mass-lifetime space for the $\tau$-neutrino (see White
et al 1995). Again, these diferences are small compared to those
induced by changes, similar to above, in $\Omega_b$ and $h$.  Finally,
as discussed above, the normalisation of the power spectrum from the
cluster abundance is uncertain by at least 15\% (1-$\sigma$)
(\cite{Eke96}). These various uncertainties limit the accuracy with
which the dark matter distribution can be calculated at the present
time.

For each cosmological model we analyse two simulations of regions of
differing size. To facilitate intercomparison, we employed the same
random number sequence to generate initial conditions for all these
simulations.  To test for finite volume effects, however, we carried
out an additional simulation of the \tCDM\ model, this time using a
different realisation of the initial conditions. In the first set of
simulations (which includes the extra \tCDM\ model), we adopted a box
length $L=239.5\hmpc$. The gravitational softening length was
initially set to 0.3 times the grid spacing and was kept constant in
comoving coordinates until it reached the value given in
Table~\ref{tbl-1}, at $z\simeq 3$. Thereafter, it was kept constant in
physical units. (The functional form of the gravitational softening
used is that given by \cite{EE81}; the values we quote correspond to
the softening scale of a Plummer potential which matches the actual
force law asymptotically at both large and small scales. The actual
force is $53.6\%$ of the full $1/r^2$ force at one softening length
and more than $99\%$ at two softening lengths.) In the second set of
simulations, the particle mass in solar masses (rather than the
volume) was kept constant in all four models and the gravitational
softening was taken to be either $30\hkpc$ or $36\hkpc$ in physical units
(after initially being kept fixed in comoving coordinates as
before). The mass resolution in these simulations is a factor of 3-20
better than in the first set. The large box simulations are large
enough to give unbiased results and relatively small sampling
fluctuations for all the statistics we study, with the exception of
large-scale bulk flows. For example, on scales $<5\hmpc$ the typical
differences in the correlation function and pair-wise velocities of
the two \tCDM\ realisations are only about 2\%. We use the large box
simulations for most of our analysis of large-scale clustering and
velocities (Sections
\ref{correlation-section}, \ref{powersp-section}, \ref{velocityfield}).  The
smaller volume simulations, on the other hand, resolve structures down to
smaller mass scales. We use these to test the effects of numerical
resolution and for a comparison with analytic models in
Section~\ref{pred-section}, where special emphasis is given to the strong
clustering regime. All our simulations have 16.7 million particles.  The
number of timesteps varied between 613 and 1588.  The \SCDM\ and \tCDM\
simulations were started at $z=50$; the \OCDM\ at $z=119$ and the
\LCDM\ at $z=30$.  The parameters of our simulations are listed in
Table~\ref{tbl-1}.

\section{Slices through the simulations\label{slices-section}}

Figures~\ref{Fig_col1}, \ref{Fig_col2}, \ref{Fig_col3} (colour plates 1, 2,
and 3) show slices through the dark matter distribution in our four models
at three different redshifts: $z=0$, 1, and 3. The slices are $239.5\hmpc$
on a side and have thickness a tenth of the side length. 
The projected mass distribution in these slices was smoothed adaptively
onto a fine grid employing a variable kernel technique similar to that
used to estimate gas densities in Smoothed Particle Hydrodynamics.

\begin{figure}
\caption{The projected mass distribution at $z=0$ in slices through four CDM
N-body simulations. The length of each slice is $239.5\hmpc$ and the
thickness is one tenth of this.  To plot these slices, the mass
distribution was first smoothed adaptively onto a fine grid using a
variable kernel technique similar to that used to estimate gas densities in
Smoothed Particle Hydrodynamics.  At $z=0$, the general appearance of all
the models is similar because, by construction, the phases of the initial
fluctuations are the same. On larger scales, the higher fluctuation
amplitude in the \LCDM\ and \OCDM\ models is manifest in sharper filaments
and larger voids compared to the \SCDM\ and \tCDM\ models. The two
$\Omega=1$ models look very similar as do the two $\Omega_0=0.3$ models
but, because of their higher normalisation, the latter show more structure.} 
\label{Fig_col1}
\end{figure}

\begin{figure}
\caption{The projected mass distribution at $z=1$ in slices through four CDM
N-body simulations. The slices show the same region as
Figure~\ref{Fig_col1}. 
The large-scale differences amongst the models are much more apparent at
$z=1$ than at $z=0$ because of the different rates at which structure grows
in these models.  The linear growth factor relative to the present value is
0.5 for \SCDM\ and \tCDM, 0.61 for \LCDM, and 0.68 for \OCDM.} 
\label{Fig_col2}
\end{figure}

\begin{figure}
\caption{The projected mass distribution at $z=3$ in slices through four CDM
N-body simulations. The slices show the same region as
Figures~\ref{Fig_col1} and~\ref{Fig_col2}.  At this early epoch 
the differences amongst the models are even more striking than at $z=1$
(c.f. Figure~\ref{Fig_col2}.) 
The linear growth factor relative to the present value is 0.25 for \SCDM\ 
and \tCDM, 0.32 for \LCDM, and 0.41 for \OCDM.}
\label{Fig_col3}
\end{figure}
\clearpage

At $z=0$, the general appearance of all the models is similar because, by
construction, the phases of the initial fluctuations are the same. The now
familiar pattern of interconnected large-scale filaments and voids is
clearly apparent. However, at the high resolution of these simulations,
individual galactic dark halos are also visible as dense clumps of a few
particles. On larger scales, the higher fluctuation amplitude in the \LCDM\
and \OCDM\ models is manifest in sharper filaments and larger voids
compared to the \SCDM\ and \tCDM\ models. Because of their higher
normalisation, the low $\Omega_0$ models also have more small-scale power
than \SCDM\ and \tCDM\ and this results in tighter virialized 
clumps. The linearly evolved power spectra of \LCDM\ and \OCDM\
are almost identical and so the primary differences between them reflect
their late time dynamics, dominated by the cosmological constant in one
case, and by curvature in the other. In \OCDM, structures of a given mass
collapse earlier and so are more compact than in \LCDM . The fine structure
in \SCDM\ and \tCDM\ is similar but since the relative amounts of
power in these models cross over at intermediate scales, clumps are
slightly fuzzier in the \tCDM\ case.

%The normalisations of the models are chosen so that the numbers of rich
%clusters matches in all the simulations thus a larger fraction of the mass
%is drawn in to make the clusters in the low $\Omega$ runs.

The large-scale differences amongst the models are much more apparent at
$z=1$. There is substantially more evolution for $\Omega=1$ than
for low-$\Omega_0$; in the former case, the linear growth factor
is 0.50 of the present value, whereas in \LCDM\ and \OCDM\ it
is 0.61 and 0.68 respectively. Thus, \OCDM\ has the most
developed large-scale structure at $z=1$, while \LCDM\ is intermediate
between this and the two $\Omega=1$ models. By $z=1$, the \OCDM\ model has
already become curvature dominated ($\Omega=0.46$) but the cosmological
constant is still relatively unimportant in the \LCDM\ model
($\Omega=0.77$).

At the earliest epoch shown, $z=3$, the differences between the models are
even more striking. The linear growth factor for \SCDM\ and \tCDM\ is
0.25 while for \LCDM\ it is 0.32 and for \OCDM\ 0.41 of its present value.
The \SCDM\ model is very smooth, with only little fine
structure. The \tCDM\ model has some embryonic large-scale structure but it
is even more featureless that \SCDM\ on the finest scales. By contrast,
structure in the low-$\Omega_0$ models, particularly \OCDM\, is already 
well developed by $z=3$.

\section{The two-point correlation functions\label{correlation-section}}

In this section we discuss the redshift evolution of the mass two-point
correlation function, $\xi(r)$, and compare the results at $z=0$ with
estimates for the observed galaxy distribution.

For each volume we have a single simulation from which to estimate
$\xi(r)$. Since this volume is assumed to be periodic, contributions
to the correlation function from long wavelength modes are poorly sampled.
In principle, it is possible to add a systematic correction, based on 
the linear theory growth of long wavelength modes (see the Appendix for
a derivation):
\begin{equation}
\bigtriangleup\xi({\bf r}) = \sum_{{\bf n}\neq(0,0,0)}^{\bf\infty}
-\xi_{lin}(|{\bf r} + L{\bf n}|)
\label{eqncorrxi}
\end{equation}
where $L$ is the simulation boxlength and $\xi_{lin}$ is the linear theory
correlation function given in terms of the linearly evolved power spectrum
$\Delta_{lin}^2$ by:
\begin{equation}
\xi_{lin}(r)=\int_0^\infty \Delta_{lin}^2\; \Big({\sin kr\over kr}\Big){dk\over k}.
\label{eqncorrflin}
\end{equation}

\begin{figure}
\plotone{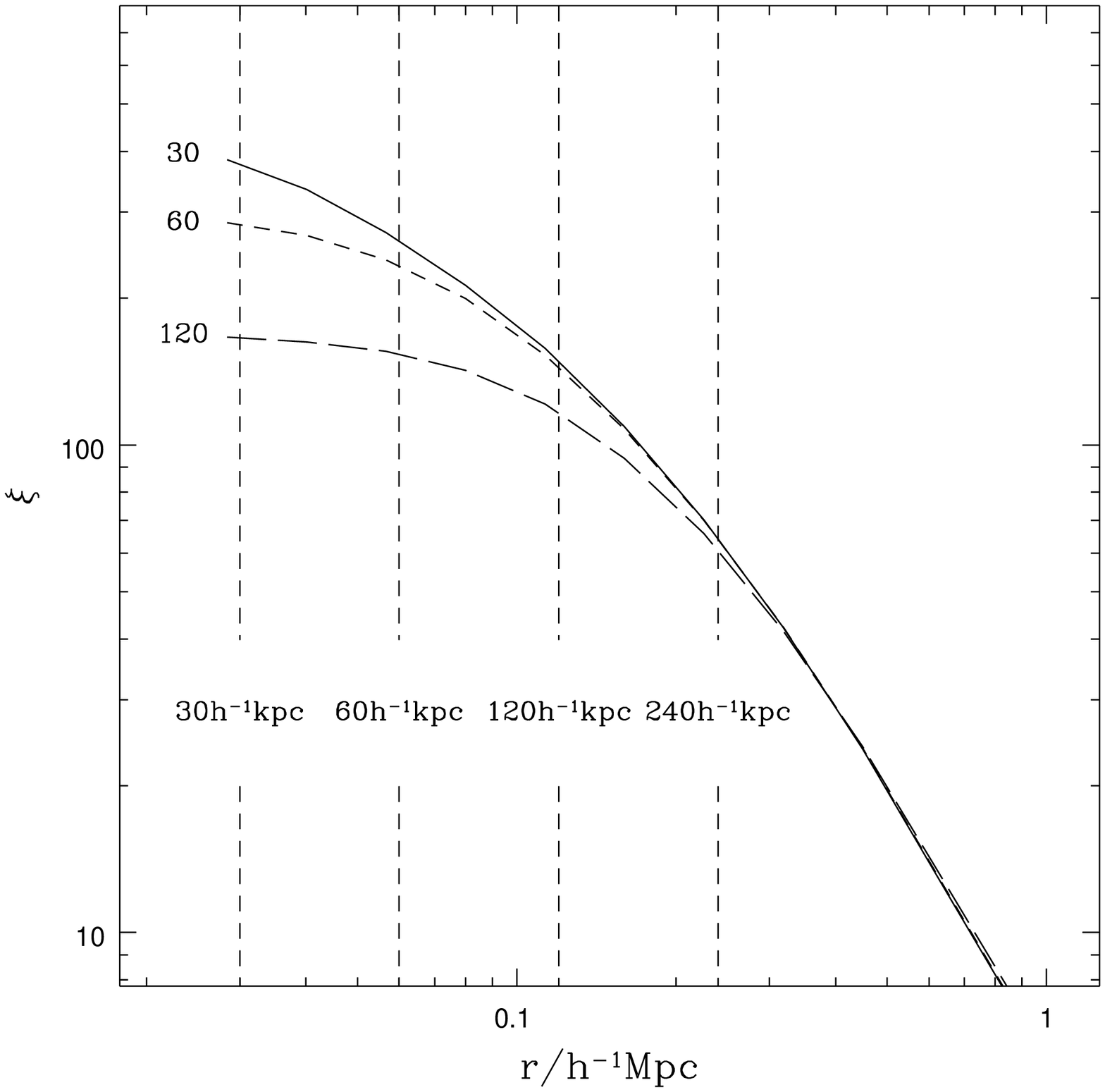}
\caption{The effect of the gravitational softening length on the two-point
correlation function. The curves show results for three $128^3$-particle
simulations of the \tCDM\ model with identical initial conditions,
but with gravitational softening lengths of 30, 60 and $120\hkpc$
respectively. Beyond twice the softening length the effect on the
correlation function is small.}
\label{fig_soft}
\end{figure}

\begin{figure}
\plotone{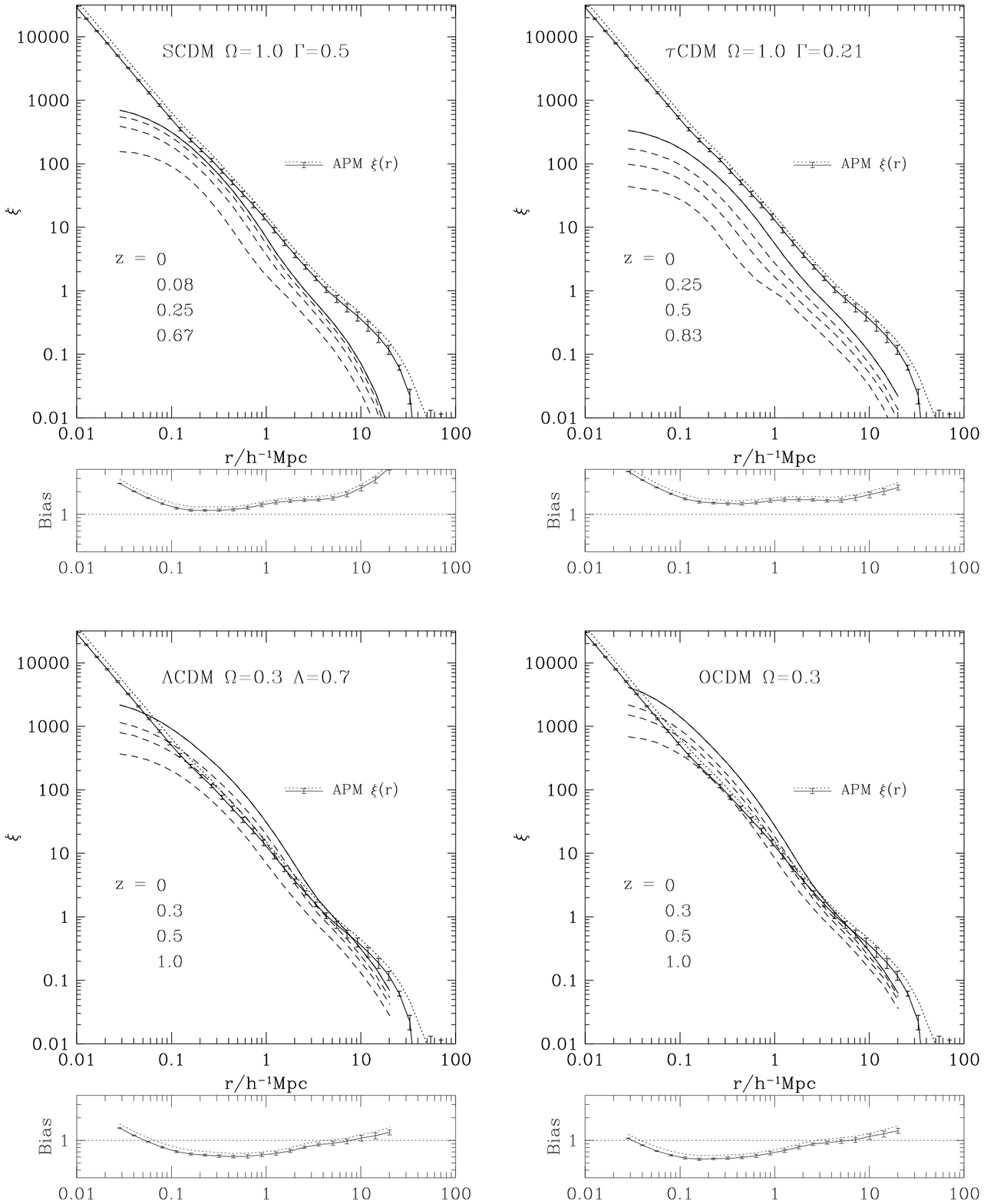}
\caption{Evolution of the mass correlation function, $\xi(r)$. The top
panels show the two-point correlation function in our four models at the
redshifts given in the legend, with results at $z=0$ plotted as a bold
solid line. The galaxy correlation function for the APM galaxy survey, 
determined by Baugh (1996), is shown as a solid line with error bars and as
a dotted line. The former corresponds to the assumption that clustering is
fixed in comoving coordinates and the latter to the assumption that
clustering evolves in proportion to the scale factor. The small panels
below each $\xi(r)$ plot show the square root of the ratio of the observed
galaxy to the theoretical mass correlation functions at $z=0$. This ratio
is the bias in the galaxy distribution that would be required for the
particular model to match the observations.}
\label{correl_fig}
\end{figure}

This expression gives a correction which is negligible for most of our
simulation volumes. For example, for \TCDMb, our simulation with the
smallest box size ($L=84.5\hmpc$) and substantial large-scale power
($\Gamma=0.21$), the correction is only 0.01 at small separations.
The expression in eqn~(\ref{eqncorrxi}) is approximately a factor of
three smaller for the $84.5\hmpc$ volume than the heuristic
correction, $\int_0^{2\pi/L}\Delta^2(\sin kr/kr)dk/k$, used by Klypin,
Primack \& Holtzman (1996). In any case, for a single simulation there
is also a random error associated with the fact that the power
originally assigned to each mode is drawn from a distribution. This
introduces a random scatter in the correlation function which is comparable
to the correction in eqn~(\ref{eqncorrxi}). The most direct way of
assessing the importance of this effect in our simulations is by comparing
two or more realizations of the same model. For the case of
\tCDM, we have carried out a second simulation with identical parameters to
the first one, but using a different random number seed to set up initial
conditions. The difference between the correlation functions of these two
simulations are less than 2\% on all scales below $<5\hmpc$, comparable to
the thickness of the line used to plot them in Figure~\ref{correl_fig}
below.

On small scales the amplitude of the two-point correlation function is
suppressed by resolution effects due to the use of softened gravity and
finite mass resolution. To test the first of these effects, we performed a
series of three simulations of the \tCDM\ model with $128^3$ particles,
identical initial conditions, the same mass resolution as the \TCDMa\
simulation, and three different values of the gravitational softening
length. The resulting two-point correlation functions are shown in
Figure~\ref{fig_soft}. The effects on the correlation function at twice the
softening length are very small. Similarly, mass resolution effects in our
simulations are small, as we discuss later in this Section and in 
Section~\ref{pred-section}.

Figure~\ref{correl_fig} shows the mass two-point correlation functions in
our four cosmological models at four different epochs. These data were
computed using the simulations \SCDMa, \TCDMa, \LCDMa, and \OCDMa. As the
clustering grows, the amplitude of the correlation function increases in a
nonlinear fashion. The overall shape of $\xi(r)$ is similar in all the
models. In all cases, ${d^2\xi/dr^2 < 0}$ on scales below $r \sim 500\hkpc$
and there is an inflection point on scales of a few megaparsecs.  The
flattening off of $\xi(r)$ at small pair separations is unlikely to be a
numerical artifact. It occurs on scales that are several times larger than
the gravitational softening length and are well resolved.  That this change
in slope is not due to mass resolution effects (associated, for example,
with the limited dynamic range of the initial conditions) is demonstrated
by the excellent agreement between the small-scale behavior of the
correlation functions plotted in Figure~\ref{correl_fig} and the
correlation functions of our smaller volume simulations which have 3-20
times better mass resolution (c.f. Figure~\ref{figure-nlxi} below; see also 
\cite{LWP91} for a discussion of why neglecting the power below the  
Nyquist frequency of the initial conditions has little effect on nonlinear
evolution.) Rather, the flattening of $\xi(r)$ at small pair separations
seems to be due to the transition into the ``stable clustering'' regime. We
return to this point in Section~\ref{pred-section}\ where we compare the 
correlation functions in the simulations with analytic models for nonlinear
evolution. 

The mass correlation functions at $z=0$ (thick solid lines) may be compared
with the observed galaxy correlation function. The largest dataset
available for this comparison is the APM galaxy survey of over $10^6$
galaxies for which Baugh (1996) has derived the two-point correlation
function, $\xi_g(r)$, by inverting the measured angular correlation function,
$w(\theta)$. The advantage of this procedure is that it gives a very
accurate estimate of the correlation function in real space, but the
disadvantage is that it requires assumptions for the redshift distribution
of the survey galaxies and for the evolution of $\xi_g(r)$ in the (relatively
small) redshift range sampled by the survey. The solid line with error bars
in Figure~\ref{correl_fig} assumes that clustering on all scales is fixed
in comoving coordinates, whilst the dotted line assumes that clustering
evolves in proportion to the scale factor. Changes in the assumed redshift
distribution produce a systematic scaling of the entire correlation
function. On scales $\gtrsim 20-30\hmpc$, the statistical error bars
may underestimate the true uncertainty in $\xi_g(r)$ since residual systematic
errors in the APM survey on these scales cannot be ruled out (\cite{Maddox96}.)

None of the model mass correlation functions match the shape of the
observed galaxy correlation function. For the galaxies, $\xi_g(r)$ is
remarkably close to a power-law over 4 orders of magnitude in amplitude
above $\xi_g=1$; at larger pair separations, it has a broad shoulder
feature. By contrast, the slope of the mass correlation functions in the
models varies systematically, so that none of the theoretical curves is
adequately fit by a single power-law over a substantial range of scales. We
have checked (Baugh, private communication) that the inversion procedure
used to derive the APM $\xi_g(r)$ from the measured $w(\theta)$ does not
artificially smooth over features that may be present in the intrinsic
clustering pattern. We have also checked that features present in the model
$\xi(r)$ are still identifiable in the corresponding $w(\theta)$ derived
with the same assumptions used in the APM analysis. The differences in shape
and amplitude between the theoretical and observed correlation functions may be
conveniently expressed as a ``bias function.'' We define the bias as the
square root of the ratio of the observed galaxy to the theoretical mass
correlation functions at $z=0$, $b(r)\equiv [\xi_g(r)/\xi(r)]^{1/2}$, 
and plot this function at the bottom of each panel
in Figure~\ref{correl_fig}. At each pair separation, $b(r)$ gives the
factor by which the galaxy distribution should be biased in order for the
particular model to match observations. For all the models considered
here the required bias varies with pair separation.

The standard CDM model, illustrated in the top left panel, shows the
well-known shortfall in clustering amplitude relative to the galaxy
distribution on scales greater than $8\hmpc$. The required bias is close to
unity on scales of $0.1-1\hmpc$, but then rises rapidly with increasing
scale. The choice of $\Gamma=0.21$ for the other models leads to mass
correlation functions with shapes that are closer to that of the galaxies
on large scales. For these models, the slope of the bias function is
relatively modest on scales $\gtrsim10\hmpc$. The large-scale behavior of
$b(r)$, however, may be affected by possible systematic errors in the APM
$w(\theta)$ at large pair separations and by finite box effects in the
simulations. The \tCDM\ model, which has the smallest amount of small scale
power, requires a significant positive bias everywhere, $b\simeq 1.5$, and
this is approximately independent of scale from $\sim 0.2-10
\hmpc$. At smaller pair separations, the bias increases rapidly. 
As discussed in the next section, the power spectrum, which is less
affected by finite box effects than the correlation function, indicates
that a constant bias for the \tCDM\ model is consistent with the APM data
even on scales larger than $10\hmpc$. Thus, uniquely amongst the models we
are considering, the shape of the correlation function and power spectrum
in the \tCDM\ model are quite similar to the observations on scales
$\gtrsim 0.2\hmpc$.

In the \LCDM\ and \OCDM\ models, the amplitude of the dark matter $\xi(r)$
is close to unity at $r=5 \hmpc$, the pair separation at which $\xi_g(r)$
is also close to unity. However, at small pair separations, the mass
correlation function has a much steeper slope than the galaxy correlation
function and, as result, $\xi(r)$ rises well above the galaxy data. Thus,
our low-density models require an ``antibias'', i.e. a bias less than
unity, on scales $\simeq 0.1-4\hmpc$. A similar conclusion was reached by 
Klypin, Primack \& Holtzman (1996) from a lower resolution N-body
simulation of a similar \LCDM\ model. As pointed out by Cole \etal (1997), 
the requirement that galaxies be less clustered than the mass must be
regarded as a negative feature of these models. Even if a plausible
physical process could be identified that would segregate galaxies and mass
in this manner, dynamical determinations of $\Omega_0$ from cluster
mass-to-light ratios tend to give values of $\Omega_0\simeq 0.2$ if the
galaxies are assumed to trace the mass (e.g. \cite{Carlberg97}). If,
instead, the galaxy distribution were actually antibiased, this argument
would result in an overestimate of the true value of $\Omega_0$. Models
with $\Omega_0$ smaller than our adopted value of 0.3, require even larger
values of $\sigma_8$, and therefore even larger antibias, in order to match
the observed abundance of galaxy clusters.  In our $\Omega=1$ models, the
required bias always remains above unity and is, in fact, quite close to
unity over a large range in scales. This is an attractive feature of these
models which may help reconcile them with virial analyses of galaxy
clusters (\cite{FEWS}), and results, in part, from the relatively low
normalisation required to match the cluster abundance. However, the bias we
infer is only about 60\% of the value required by Frenk \etal (1990) to
obtain acceptable cluster mass-to-light ratios in an $\Omega=1$ CDM
cosmology with ``high peak'' biasing.

It seems almost inevitable that the process of galaxy formation and
subsequent dynamical evolution will bias the galaxy distribution relative
to the mass in a complicated way. Indeed, a variety of biasing mechanisms
have been discussed in the past. These are essentially of two types.  In
the first, galaxy formation is assumed to be modulated, for example, by the
local value of the density smoothed on cluster scales, as in the high
peak bias model of galaxy formation (\cite{BBKS}; DEFW), or by the
effects of a previous generation of protogalaxies (e.g. \cite{DR87}). Such
local processes tend to imprint features on the galaxy correlation function
on small and intermediate scales, but Coles (1993) and Weinberg (1995) have
argued that they do not appreciably distort the shape of the mass
correlation function on large scales. This, however, may be achieved by
some form of non-local bias like in the ``cooperative galaxy formation''
scheme proposed by Bower \etal (1993; see also \cite{BB91}). In this case,
a match to the APM $w(\theta)$ on large scales is possible with a suitable
choice of model parameters. The second type of biasing mechanism is of
dynamical origin. An example is the ``natural bias'' found in the CDM
simulations of White \etal (1987b) who showed that the dependence of
fluctuation growth rate on mean density naturally biases the distribution
of massive dark halos towards high density regions (see also \cite{Cen92}.)
Another example is dynamical friction which, as Richstone, Loeb \& Turner
(1992) and Frenk \etal (1996) amongst others have shown, can segregate 
galaxies from mass in rich clusters. Dynamical biases of this type tend to
enhance the pair count at small separations, flattening the bias function
on scales of a few hundred kiloparsecs. Mergers, on the other hand, have
the opposite effect and may even give rise to an antibias of the kind
required in our low-$\Omega_0$ models (c.f. \cite{Jenk97}). Thus, it seems
likely that the correlation function of the {\it galaxies} that would form
in our models will differ from the correlation function of the
mass. Nevertheless, the fine tuning required to end up with an almost
featureless power-law correlation function over at least two orders of
magnitude in scale seems a considerable challenge for this general class of
models.

\begin{figure}
\plotone{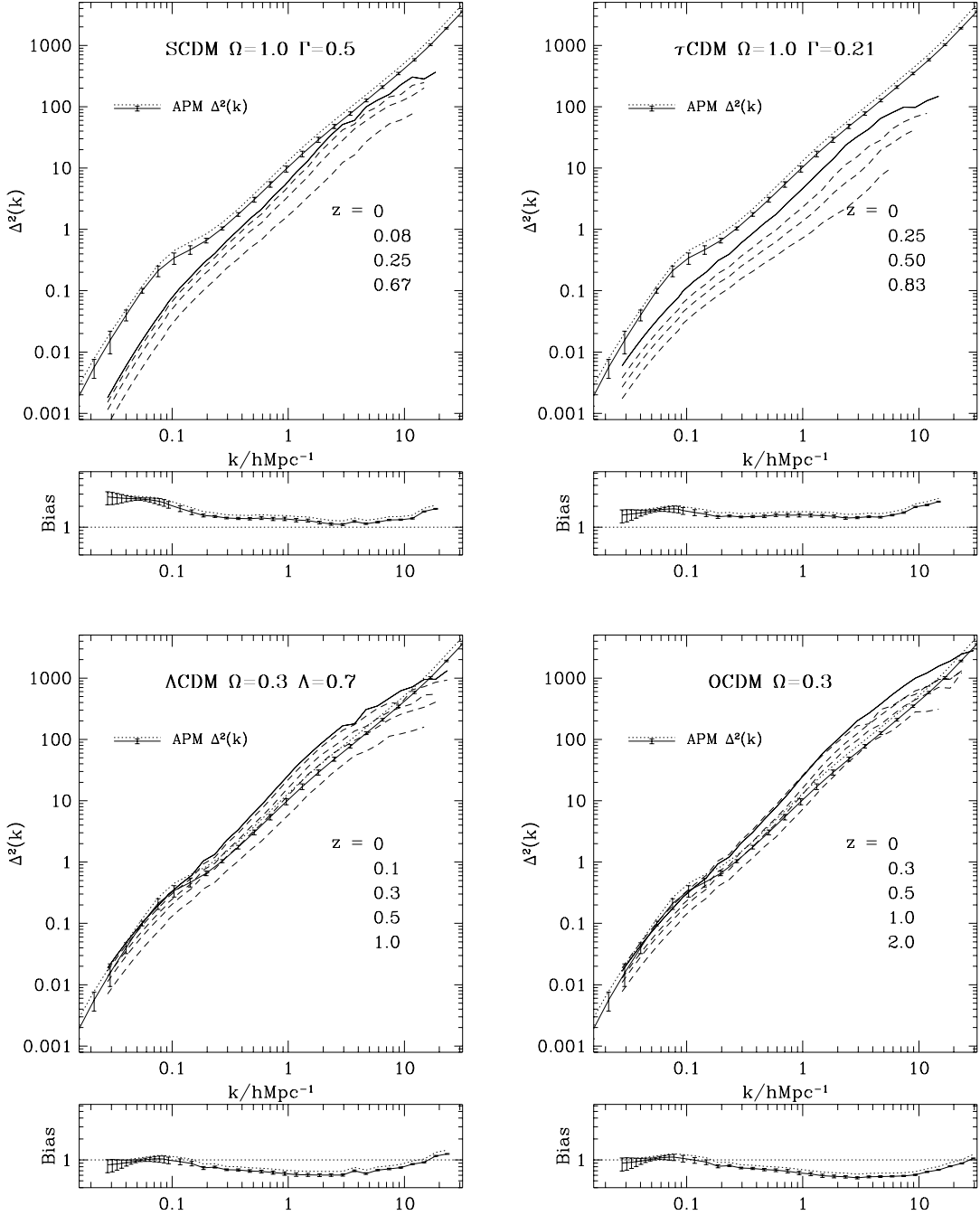}
\caption{ Evolution of the power spectrum of the dark matter in the
simulations. The large panels show the power spectrum evaluated at the 
redshifts given in the figure legend, with results at $z=0$ shown as a 
solid line.  The solid line with error bars and the dotted line are
estimates of the power spectrum of the APM galaxy survey obtained assuming,
respectively, that clustering is fixed in comoving coordinates or that it
grows with the scale factor (Baugh \& Efstathiou 1993). The small panels
show the square root of the ratio of the APM galaxy power spectrum to the
present day dark matter spectrum. This ratio is the bias in the galaxy
distribution required for the model to match the APM data. For 
$k<0.086 h/{\rm Mpc}$ the linear theory power spectrum has been plotted, 
rather than the actual spectrum which is noisy due to the small number of 
modes that contribute to each bin.}
\label{power_figure}
\end{figure}

\section{The power spectra\label{powersp-section}}

For an isotropic distribution in $k$-space, the power spectrum is related
to the correlation function by
\begin{equation}
\xi(r)=\int_0^\infty \Delta^2(k)\;\Big({\sin kr\over kr}\Big) {dk\over k}. 
\label{eqncorrf}
\end{equation} 

To measure the power spectrum of our simulations over a wide range of
scales we use a technique which is efficient both in terms of computational
expense and memory. To evaluate the power spectrum on the smallest scales,
we divide the computational volume into $m^3$ equal cubical cells and
superpose the particle distributions of all $m^3$ cells. The Fourier
transform of this density distribution, which is now periodic on a scale
$L/m$, recovers exactly the power present in the full simulation volume in
modes which are periodic on the scale $L/m$. These modes form a regular
grid of spacing $2m\pi/L$ in ${\bf k}$-space. The estimate of $\Delta^2(r)$
is obtained by averaging the power of large numbers of modes in spherical
shells. Provided these modes have, on average, representative power this
gives an unbiased estimate of the power spectrum of the simulation.  In
principle, the power of all the modes in the full simulation can be
obtained by applying a complex weighting, $\exp(2\pi i{\bf n\cdot r}/L)$, to
a particle at position ${\bf r}$ during the charge assignment prior to
taking the discrete fast Fourier transform. This charge assignment creates
a uniform translation in ${\bf k}$-space by $2\pi{\bf n}/L$. With a
suitable choice of ${\bf n}$ one can recover a different set of modes from
the original simulation, always with a spacing of $2m\pi/L$ in ${\bf
k}$-space. Applying this method $m^3$ times allows the recovery of all
modes present in the simulation, although there is no longer any gain in CPU
time over a single large fast Fourier transform. Because of the 
sparse sampling of ${\bf k}$-space, the estimate of the power on the
scale $L/m$ has a large variance. However, by using a $64^3$ mesh and
evaluating the Fourier transform for several values of $m$ one can
evaluate the power spectrum on any scale with adequate sampling and avoid
this problem except for $m=1$.

The assumption that these sparsely sampled modes carry representative power
is true by construction in the initial conditions. The violation of this
assumption as a result of nonlinear evolution is very unlikely because it
would require a detailed large-scale ordering to develop over the
simulation. This may, however, come about artificially; for example, the
MAPS procedure of Tormen and Bertschinger (1996, see also \cite{Cole97b}), which
is designed to extend the dynamic range of an N-body cosmological
simulation, requires periodically replicating a simulation and then
modifying the large-scale modes so as to effectively add large-scale power
not present in the original simulation. In this case, the large-scale order
arising by the replication introduces significant fine scale structure in
$k$-space (\cite{Cole97b}) and one should be wary when applying this method.

Figure~\ref{power_figure} shows the time evolution of the power spectrum
for the same four simulations ($L=239.5 \hmpc$) illustrated in
Figure~\ref{correl_fig}. As before, two graphs are shown for each
model. The larger one gives the time evolution of the power spectrum,
plotted at four different epochs. The $z=0$ results may be compared with
the 3D power spectrum of the APM galaxy survey (\cite{BE93}). As for the
correlation function, two versions of the APM power spectrum are plotted,
one assuming that the clustering pattern remains fixed in comoving
coordinates (solid curve with error bars) and the other assuming that it
evolves in proportion to the scale factor (dotted curve). For wavenumbers
$k<0.086h/{\rm Mpc}$ we have plotted the linear theory power spectrum rather
than the simulation results since the sparse sampling of the modes with
wavelength comparable to the simulation box size gives rise to spurious
fluctuations. The linear extrapolation can be seen to join smoothly onto
the actual power spectrum on these scales. The smaller panels show the
square root of the ratio of the APM galaxy power spectrum to that of the
dark matter in the simulation at $z=0$. As before, this is the
scale-dependent bias required in the galaxy distribution for a
particular model to be a good match to the APM data.

Comparison of the APM data with the power spectrum of the dark matter
in the different cosmological models brings out essentially the same
features as the corresponding comparison with the correlation function. In
the SCDM model, the dark matter power spectrum falls below that of the
galaxies at small wavenumbers, requiring a bias function that increases
rapidly at small $k$. The shape of the power spectrum in the low-$\Omega_0$
models is similar to that of the APM galaxies only for $k < 0.1h/{\rm
Mpc}$; at larger $k$ the dark matter distribution has more power than the
galaxy distribution, requiring a bias less than unity. Only the \tCDM\
model has a dark matter power spectrum whose shape matches that of the
galaxy data over a wide range of scales. The required bias in this case is
approximately constant for $0.02 \lesssim k/h{\rm Mpc}^{-1}\lesssim 10$.

\section{Comparison with analytic predictions} 
\label{pred-section}

We now compare the results of our simulations with a 
parameterised fitting formula which Peacock \& Dodds (1996) use to 
predict the power spectrum of the nonlinear mass density field which
develops through gravitational amplification of any given gaussian field
of linear density fluctuations. We consider both the power
spectrum and the correlation function. We first
summarise the theory and then compare it with the simulation results
discussed in Sections~\ref{correlation-section} and~\ref{powersp-section}.

\subsection{Method}

Hamilton \etal (1991) suggested a formalism for computing the
nonlinear growth of the two-point correlation function.  Peacock \&
Dodds (1994) adapted this method to the computation of nonlinear power
spectra, and extended it to cosmologies with $\Omega_0\ne 1$. 
Baugh \& Gaztanaga (1996) applied it to the power spectrum of the APM
galaxy survey. The original formalism of Hamilton \etal (1991) was
independent of the shape of the power spectrum, but Jain, Mo \& White
(1995) showed that this is not correct. Peacock \& Dodds (1996) give
an improved version of the Peacock \& Dodds (1994) method which takes
this into account and allows the nonlinear spectrum produced by
evolution from any smoothly-varying linear spectrum to be
calculated. \cite{Smith97} have tested the new procedure with a large
number of N-body simulations.  The method may be summarized as follows.

The nonlinear spectrum is a function of the linear spectrum at a smaller
linear wavenumber:
\begin{equation}
\Delta^2_{\ts NL}(k_{\ts NL}) = f_{\ts NL}[\Delta^2_{L}(k_{\ts L})], 
\label{delta_nl}
\end{equation}
\begin{equation}
k_{L} = [1+\Delta^2_{\ts NL}(k_{\ts NL})]^{-1/3} k_{\ts NL}. 
\end{equation}
The following fitting formula for the nonlinear function, $f_{\ts NL}$ 
was proposed by Peacock \& Dodds (1996): 
\begin{equation}
f_{\ts NL}(x) =x \; \left[{ 1+B\beta x +[A x]^{\alpha\beta} \over
1 + ([A x]^\alpha g^3(\Omega_0)/[V x^{1/2}])^\beta}\right]^{1/\beta}.
\end{equation}
In this expression, $B$ describes a second-order deviation from linear
growth; $A$ and $\alpha$ parametrise the power-law which dominates the
function in the quasi-linear regime; $V$ is the virialisation parameter
which gives the amplitude of the $f_{\ts NL}(x) \propto x^{3/2}$ asymptote
(where the behaviour enters the ``stable clustering'' limit); and $\beta$ softens
the transition between these regimes. For power spectra of the form 
$|\delta_k^2|\propto k^n$, the parameters and their dependence on $n$ are: 
\begin{eqnarray}
A & = &0.482\,(1+n/3)^{-0.947} \nonumber \\
B & = &0.226\,(1+n/3)^{-1.778} \nonumber \\
\alpha & = &3.310\,(1+n/3)^{-0.244}\\
\beta & = & 0.862\,(1+n/3)^{-0.287}\nonumber\\
V & = &11.55\,(1+n/3)^{-0.423}.\nonumber\\
\nonumber \end{eqnarray}
The growth factor, $g(\Omega)$, is proportional to the ratio of the linear
growth factor to the expansion factor. It takes the value unity for
$\Omega=1$ and, for $\Omega_0<1$, it tends to unity as $\Omega\rightarrow
1$.

\begin{figure}
\plotone{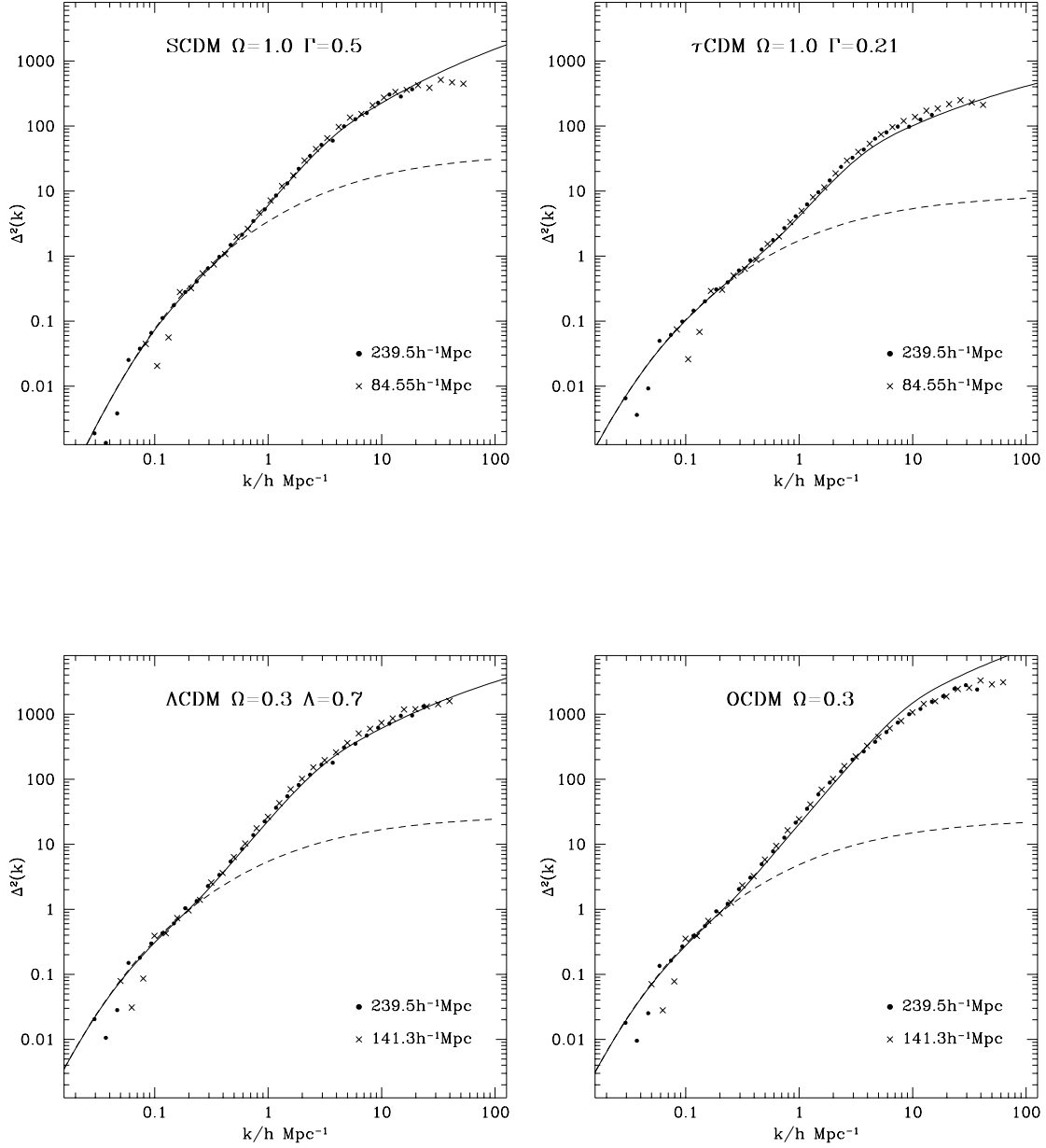}
\caption{Predicted nonlinear power spectra at $z=0$ compared with N-body
simulation results. The analytical results for our four cosmological models
are shown as solid curves and the N-body results in our large and small
volume simulations are shown by solid dots and crosses respectively. The
dashed line shows the linear theory prediction for the power spectrum at
$z=0$. At small wavenumbers the simulations depart from the linear theory
curve because of the small number of modes in each bin.}
\label{figure-nlpow}
\end{figure}
\begin{figure}
\plotone{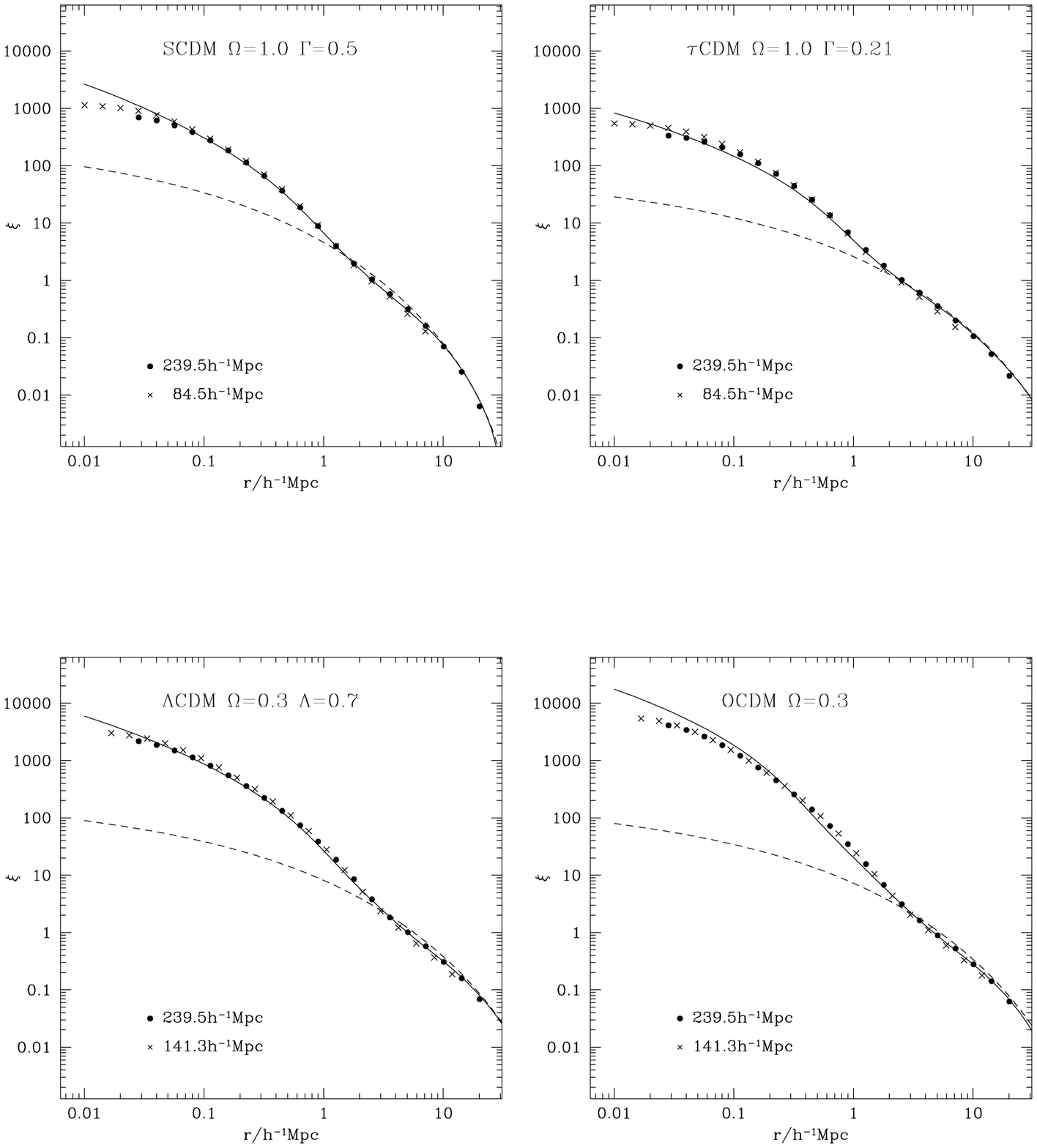}
\caption{Predicted mass correlation functions at $z=0$ compared with N-body
simulation results. The analytical results for our four cosmological models
are shown as solid curves and the N-body results in our large and small
volume simulations are shown by solid dots and crosses respectively. The
dashed line shows the linear theory prediction for $\xi(r)$ at 
$z=0$. At large pair separations the integral constraint in the smaller
simulations depresses $\xi(r)$ slightly, whereas at small pair
separations, $\xi(r)$ is slightly higher in the smaller volumes because
they have better mass resolution.}
\label{figure-nlxi}
\end{figure}

For linear spectra which are not a power-law, particularly for the CDM
model, Peacock \& Dodds (1996) suggested that a tangent spectral index as a 
function of linear wavenumber should be used:
\begin{equation}
n_{\rm eff}(k_{\ts L})\equiv {d\ln P \over d \ln k}(k=k_{\ts L}/2).
\label{neff}
\end{equation} 
The factor of 2 shift to smaller $k$ is required because the tangent
power-law at $k_{\ts L}$ overestimates the total degree of nonlinearity
for curved spectra in which $n_{\rm eff}$ is a decreasing function of $k$
and underestimates it in the opposite case. Peacock \& Dodds (1996) 
state that this 
prescription is able to predict the nonlinear evolution of power-law and
CDM spectra up to $\Delta^2\simeq 10^3$ with an rms precision of about
7\%. Since the fitting formula is designed to reproduce the results for
power-law spectra, the main uncertainty in this method is whether or not the
shifted tangent power-law is the best means of deducing the effective $n$
as a function of scale. This issue becomes especially important when the
effective index is more negative than $-2$ (because nonlinear effects
diverge as $n\rightarrow -3$), and when the curvature of the spectrum is
especially severe. This means that spectra with low values of $\Omega_0 h$
or of $\sigma_8$ present the greatest challenge for the analytic
method.

The effect of cosmology enters into the fitting formula only through the
growth factor, $g(\Omega)$, which governs the amplitude of the virialised 
portion of the spectrum.

%According to \cite{CPT}, the growth factor may be
%approximated almost exactly by
%\begin{equation}
%g(\Omega_0) ={{5}\over{2}}\Omega_{\rm m}\left[\Omega_{\rm m}^{4/7}-\Omega_{\rm v}+
%(1+\Omega_{\rm m}/2)(1+\Omega_{\rm v}/70)\right]^{-1},
%\end{equation}
%where matter ($m$) and vacuum ($v$) contributions to the density parameter
%are distinguished; $\Omega$ without a subscript generally means 
%$\Omega_{\rm m}$.
%When considering non-zero vacuum energy, it is usual to restrict
%attention to spatially flat models only. Models with
%$\Omega<1$ thus come in two varieties: open ($\Omega_v=0$)
%and flat ($\Omega_v=1-\Omega_m$).

\subsection{Fit to the simulations.\label{evol-pow-xi-fit}}

The nonlinear power spectrum predicted by eqn~(11) for each of our four
cosmological models is plotted as a solid line in
Figure~\ref{figure-nlpow}. The solid circles and crosses show the results
from our large and small volume simulations respectively. Note the
excellent agreement between them. The dashed curve shows the linear theory
prediction for the present day power spectrum\footnote{The realisation of
the power spectrum in our simulations can be seen to have a
downward fluctuation in power at $1\le |kL/2\pi| < 2$, where $L$ is the
simulation box size. A $\chi^2$ test for these 26 modes shows that a
fluctuation lower than this is expect in 7\% of cases. While this
fluctuation is not particularly unusual, it has little effect on the
results of interest (except for bulk flows; c.f. \S3.1, \S5 and \S8)
because our simulated volumes are sufficiently large.}.  The points are plotted only on
scales where the power exceeds the shot noise. The agreement between the
analytical and numerical results is generally good, particularly for \SCDM\
and \LCDM. For all the models with $\Gamma=0.21$, the predicted power
spectrum slightly underestimates the detailed power spectrum of the
simulations around the region $\Delta^2\simeq10$. As discussed above, these
cases are expected to be especially challenging, because they have a more
negative $n_{\rm eff}$ at the nonlinear scale. The slight mismatch
illustrates the difficulty in defining precisely the effective power-law
index for these rather flat spectra, and a more accurate formula could be
produced for this particular case, if required. Note that in the
quasilinear portion the power spectra follow very closely the general shape
predicted by eqns~(\ref{delta_nl})-(\ref{neff}); in particular, there is
essentially no difference between the \OCDM\ and \LCDM\ results, as
expected.

The power spectra of the different cosmological models are expected to part
company at higher fequencies, where the spectrum enters the ``stable
clustering'' regime, and indeed they do. However, although the
predictions match the \LCDM\ results almost precisely at
$\Delta^2\simeq 1000$, they lie above the \OCDM\ results at high $k$:
$\Delta^2(k=30) \simeq 4500$, compared to the simulation value of
2500. At one level, this is not so surprising, since the smaller
simulations that Peacock \& Dodds (1996) used to derive the parameters
of the fitting formula were not able to resolve scales beyond
$\Delta^2\simeq 1000$.  However, the amplitude of the stable
clustering asymptote is very much as expected in the $\Omega=1$ and
\LCDM\ cases, and the argument for how this amplitude should scale
with $\Omega_0$ is straightforward: at high redshift, clustering in
all models evolves as in an $\Omega=1$ universe, and so evolution to
the present is determined by the balance between the linear growth
rate and the ($\Omega_0$ independent) rate of growth of stable
clustering. The failure of this scaling for the \OCDM\ case is
therefore something of a puzzle. It is conceivable that the numerical
result could be inaccurate, since it depends on resolving small groups
of particles with overdensities of several thousand, and these
collapse very early on. However, we have verified that changing the
starting redshift from 59 to 119 does not alter the results of the simulations
significantly.
 
Figure~\ref{figure-nlxi}\ shows the two-point correlation function derived
using eqn~(\ref{eqncorrf}) and the predicted nonlinear power spectrum,
eqns~(\ref{delta_nl})-(\ref{neff}). As before, the N-body results are
plotted as filled circles and crosses for the large and small volume
simulations respectively. Note that in general, the agreement between each
pair of simulations is very good and the very small discrepancies that
there are can be understood simply. At large pair separations $\xi(r)$ is
slightly depressed in the smaller simulations because these separations are
becoming an appreciable fraction of the box length and the integral
constraint requires $\xi(r)$ to average to zero over the volume of the
simulation.  At small pair separations, $\xi(r)$ is slightly higher in the
smaller volumes because of their higher mass resolution. Once again, there
is good agreement in general between the anlytical predictions and the
N-body results, particularly for the \LCDM\ and \SCDM\ models. For \tCDM,
the model underpredicts the correlation function on scales below
$~700\hkpc$ whilst for \OCDM, the model correlation function is somewhat
steeper than in the simulations. These differences occur on scales
significantly larger than those affected by resolution effects, and are
fully consistent with the analogous deviations seen in the power spectrum.

\section{The Velocity Fields and distributions.\label{velocityfield}}

In this section we compute bulk flows, velocity dispersions, and pairwise
velocities of the dark matter particles in our simulations. Potentially,
measurements of galaxy peculiar velocities can provide powerful tests of
the models. In practice, there are a number of complications which weaken
these tests. Foremost amongst them is the uncertain relation between the
velocity fields of dark matter and galaxies, particularly on small scales
where various dynamical biases may operate (\cite{Carlberg90}, \cite{FEWS}). 
It is 
relatively straightforward to calculate, with high precision, the velocity
fields of the dark matter in a given cosmology, using simulations like ours
or, in the appropriate regime, using linear theory. To relate these to
observations on small scales requires an understanding of possible
dynamical biases and, in the case of pair-weighted statistics, of sampling
uncertainties and systematic effects arising from the discrete nature of
the galaxy population. Only on sufficiently large scales do we expect
galaxy bulk flows which are, in principle, measurable to be simply related
to the dark matter bulk flows.

Observational determinations of galaxy velocities have their own
complications. For example, determining bulk flows over
representative volumes requires measuring peculiar velocities, and thus
determining distances with an accuracy of a few percent, for large samples of
galaxies. Defining such samples in a homogeneous way and keeping systematic
effects in the distance measurements within tolerable levels is a complex
and still uncertain process (e.g. \cite{Willick97}).  Other measures of the
galaxy velocity field such as the pairwise relative velocities of close
pairs are also affected by systematic and sampling effects even though they
do not require measuring distances (e.g. \cite{Mar95}; \cite{MJB96}.) 

In view of the various uncertainties just mentioned, we focus here on high
precision estimates of various measures of the dark matter velocity
field. Our main purpose is to contrast the velocity fields predicted in the
four cosmological models considered in this paper, in the expectation that
these and related calculations may eventually be applied to a reliable
interpretation of real galaxy velocity fields.  We do, however, carry out a
limited comparison of dark matter velocity fields with existing data on
large-scale galaxy bulk flows and pairwise velocity dispersions. In 
subsection~\ref{bulkflow}\ we compute
distributions of the mean and rms dark matter velocity on various scales
and in subsection \ref{velco-section}\ we consider pairwise velocities also
over a range of scales.

\subsection{Bulk flows and dispersions.\label{bulkflow}}

\begin{figure}
\plotone{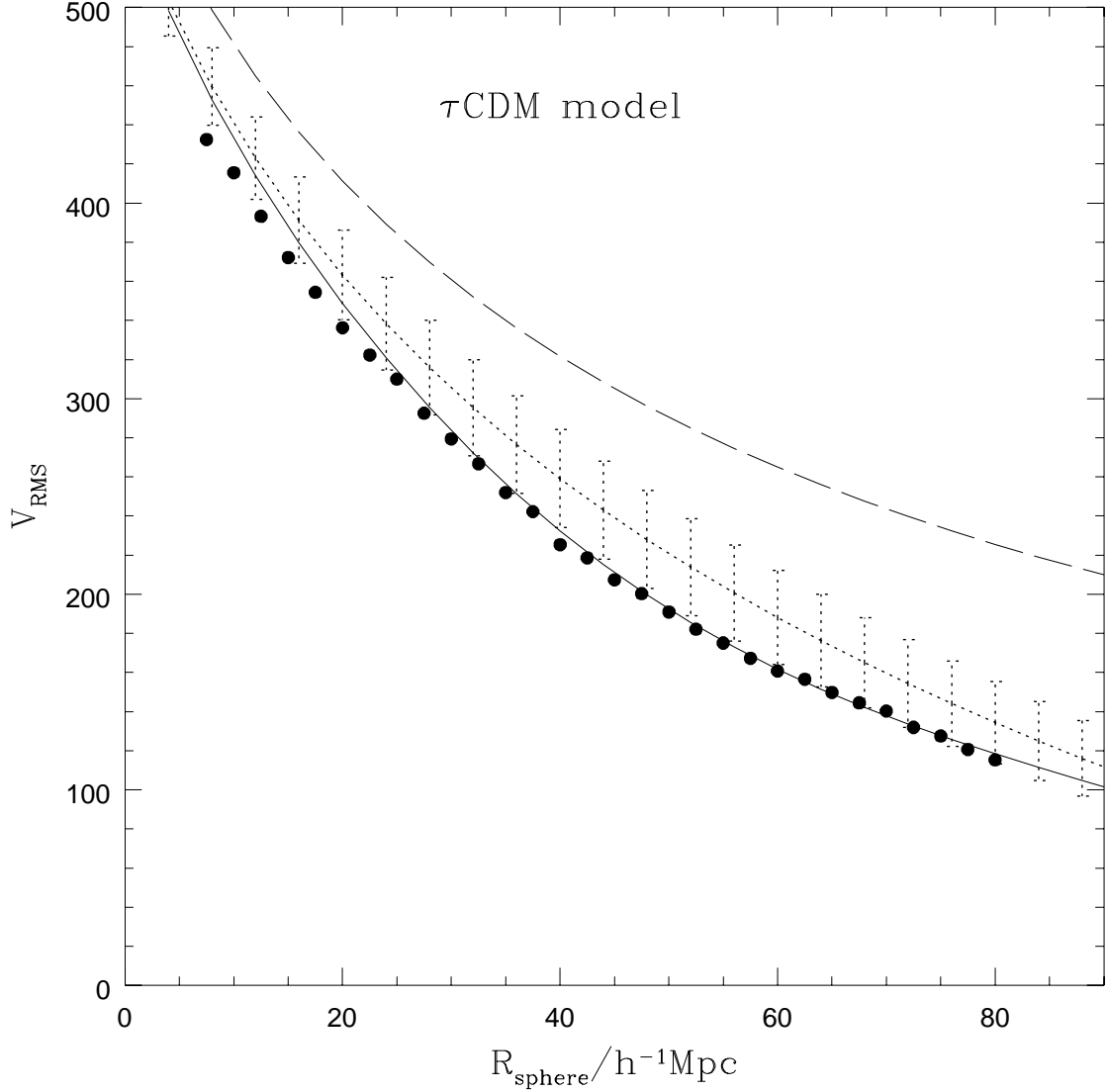}
\caption{Comparison of the bulk flow measured in the \tCDM\ model (solid
circles)
with linear theory. The long-dashed curve is the linear theory result in
the limit of an infinite box size. The dotted line with error bars shows
the ensemble rms average for a $239.5\hmpc$ periodic box.  The error bars
give the rms spread between different realisations. The solid line is the
result from linear theory for the realisation used in our \tCDM\
simulation. Linear theory works to excellent approximation when all the
finite box effects are taken into account.}
\label{tcdmlin}
\end{figure}

We compute bulk flows and velocity dispersions of dark matter particles in
the simulations by placing a large number of spheres of varying radii
around random locations in the computational volume. We define the bulk
velocity of a sphere as:
\begin{equation}
    {\bf V} = {1\over N}\sum_{i=1,N} {\bf v_i}
\end{equation}
where ${\bf v_i}$ is the peculiar velocity of the $i$th particle out of $N$
in a given sphere and all particles have equal weight. The dispersion
$\sigma_{v}$ is defined as:
\begin{equation}
    \sigma^2_{v} = {1\over {N-1}}\sum_{i=1,N} ({\bf v_i - V})^2
\end{equation}

In linear theory, the bulk velocity of the dark matter can be 
accurately calculated according to:
\begin{equation}
    <V^2> = \Omega_0^{1.2}\int_0^{\infty} k^{-2}W^2(Rk)\Delta^2(k){{\rm d}k\over k}
\label{vel_int}
\end{equation}
where $W(Rk)$ is a window function, which we take to be a top hat of radius
$R$ in real space. The approximate factor $\Omega_0^{1.2}$ works well
for all the cosmological models we are considering here (\cite{PJEP80}.) 

The integral in eqn~(\ref{vel_int}) ranges over all spatial scales and so
applies to a simulation only in the limit of an infinite volume. In order
to compare the simulations with linear theory it is necessary to take
account of effects due to the finite computational box and of the fact that
we have only one realisation. Finite box effects are much more significant
for velocities than for the correlation function (eqn~\ref{eqncorrflin}),
since the relative importance of longer waves is enhanced in
eqn~(\ref{vel_int}) by a factor $k^{-2}$. To compare linear theory with a
specific simulation, the integral in expression~(\ref{vel_int}) must be
replaced by a summation over the modes of the periodic box, using the
appropriate power in each mode as set up in the initial conditions.

The dashed curve in Fig~\ref{tcdmlin} shows the linear theory prediction
for bulk flows at $z=0$, in spheres of radius $R_{sphere}$, for a model with
the power spectrum and normalisation of our \tCDM\ simulation, in the limit
of infinite volume. The predicted velocities fall off smoothly from about
$500\,\kms$ at $10 \hmpc$ to about $200\,\kms$ at $100 \hmpc$.  The dotted
curve shows the linear theory ensemble average value of $<V^2>^{1/2}$ over
realizations of the \tCDM\ power spectrum in volumes the size of our
simulation. The difference between this and the dashed curve indicates just
how important finite box effects are in computing bulk flows.  The error
bars on the dotted curve show the rms dispersion amongst different
realizations. For small spheres, the variation about the mean is
approximately Gaussian and the error bars may be regarded as 1-$\sigma$
deviations from the mean. The results from our actual simulation at $z=0$
are plotted as solid circles in the figure and the linear theory prediction
for evolution from the specific initial conditions of this simulation is 
shown as the solid curve. The particular realisation that we have simulated
turned out to produce slightly, but not anomalously, low velocities. On
scales above $20\hmpc$ the linear theory prediction agrees very well with
the simulation; at $R=10\hmpc$, it overestimates the actual velocities by
$5\%$. 

While linear theory suffices to calculate bulk flows on scales larger that
about $10\hmpc$, the velocity dispersion of particles in spheres is
dominated by contributions from nonlinear scales and must be obtained from
the simulations. Finite box effects are not important in this case because
the contributions from wavelengths larger than the simulation box are small.

\begin{figure}
\plotone{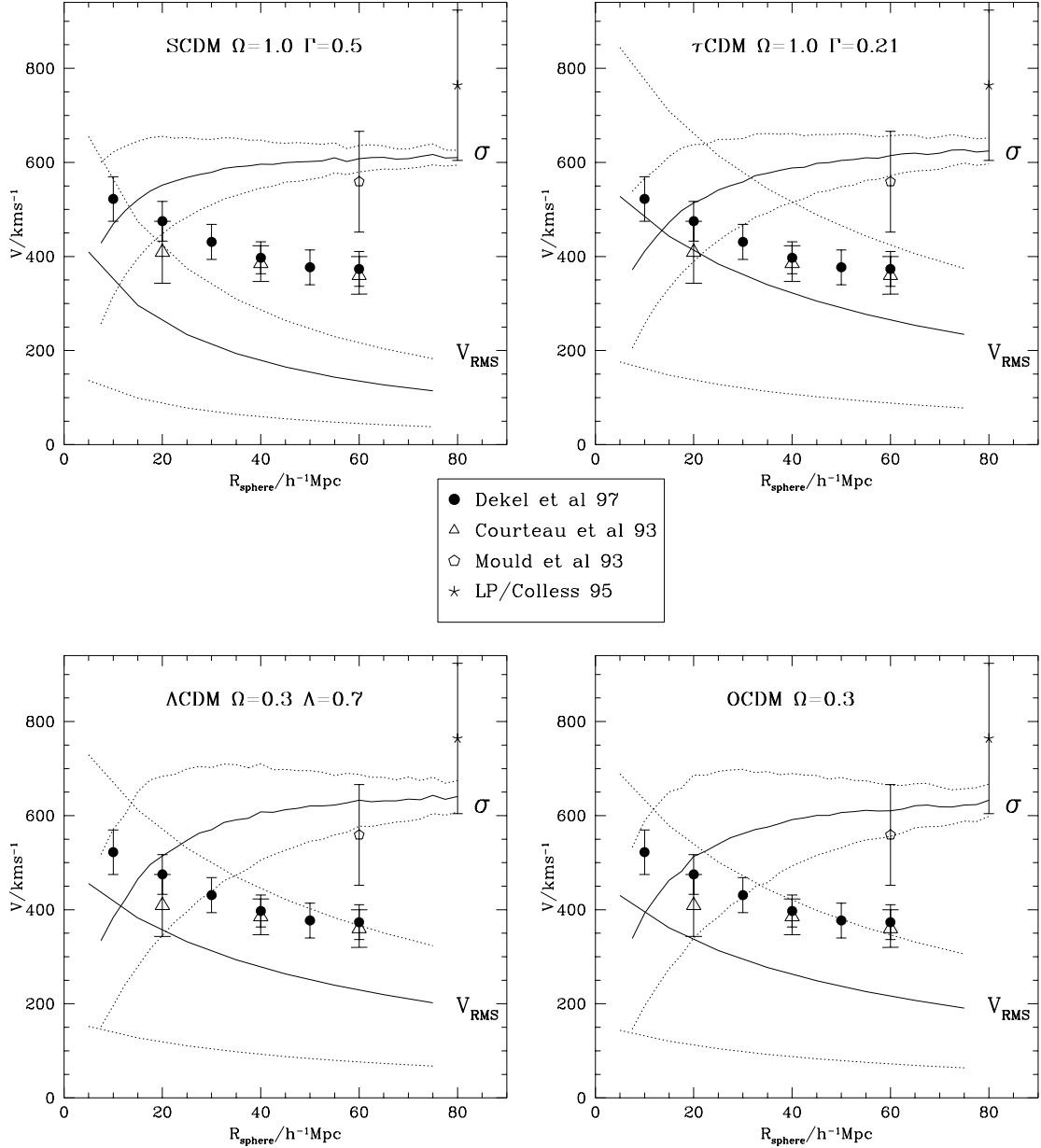}
\caption{Dark matter bulk flows and velocity dispersions in spheres of 
different radii. The bulk flows, computed from linear theory, are
shown by the lower solid line, with 90\% confidence limits indicated
by dotted lines. The rms velocity dispersions, computed from the
simulations, are shown by the upper solid curve, with the rms scatter
indicated by the dotted lines. The data points with error bars are
observational estimates of galaxy bulk flows from Dekel \etal\ (1997),
Courteau \etal\ (1993), Mould \etal\ (1993), and Lauer and Postman
(1993), as reanalysed by Colless (1995). (See legend
in the middle of the Figure.) The predicted velocity
fields are very similar in all the models because they are normalised
to give the same abundance of rich clusters. The only exception are
the predicted bulk flows in the \SCDM\ model which are slightly
smaller than in the other models because of its different power
spectrum shape. Every model except \SCDM\ is consistent with the
galaxy bulk flow data with the exception of the Lauer and Postman
result.}
\label{bulkfig}
\end{figure}

The bulk flows, $<V^2>^{1/2}$, calculated from linear theory and the
velocity dispersions in spheres, $\sigma_v$, calculated from our
$L=239.5\hmpc$ simulations are plotted as solid lines in Fig~\ref{bulkfig}
for our four cosmological models. The dotted curves around the
$<V^2>^{1/2}$ curve correspond to $90\%$ confidence limits on the bulk
velocity for a randomly placed sphere, calculated by integrating over the
appropriate Raleigh distribution. 
The dotted curves around the
$\sigma_v$ curve indicate the rms scatter of the $\sigma_v$ distribution.

With the exception of \SCDM\ the predicted bulk flows in all our models are
remarkably similar. The reason for this can be traced back to our choice of
normalisation which ensures that all models produce approximately the same
number of rich galaxy clusters. This choice effectively cancels out the
dependence of the bulk flow velocity on $\Omega_0$ as may be seen directly
from linear theory.  From eqn~(\ref{vel_int}), $<V^2>^{1/2}\propto
\sigma_8\Omega_0^{0.6}$, for a fixed shape of the power spectrum. On the
other hand, our adopted fluctuation normalisation requires approximately
that $\sigma_8\propto \Omega_0^{-0.5}$ (cf. eqns~1 and~2). Since the power
spectra of the \LCDM, \tCDM, and \OCDM\ models all have the same shape
parameter, $\Gamma=0.21$, the bulk flows in these models are very
similar. The lower bulk flow velocities predicted in the \SCDM\
model reflect the relatively smaller amount of large scale power in this
model implied by its value of $\Gamma = 0.5$. The mean bulk velocity in
\SCDM\ is approximately 2/3 of the value in the other models.

The peculiar velocity dispersion of dark matter particles in random
spheres is also remarkably similar in all our models, including \SCDM. In
this case, significant contributions to $\sigma_v$ come from a wide range
of scales, including nonlinear objects as well as regions which are still
in the linear regime. On small scales, $\sigma_v$ rises with increasing
sphere radius and reaches a plateau at radii of a few tens of megaparsecs.
The limit as the radius tends to infinity is just the single particle rms
peculiar velocity. For our large simulation boxes, this is $614\,\kms$,
$635\,\kms$, $648\,\kms$\ and $630\,\kms$ for the \SCDM, \tCDM, \LCDM\ and
\OCDM\ models respectively. The slightly lower value for \SCDM\ again
reflects the smaller large-scale power in this model compared to the
others. This deficit on large scales, however, is compensated by an excess
contribution from smaller scales.

We have plotted in Figure~\ref{bulkfig}\ estimates of galaxy bulk flow
velocities in the local universe taken from the analyses by Mould \etal
(1993), Courteau \etal (1993), Dekel \etal\ (1997), and Lauer \& Postman
(1994). These estimates are based on different datasets and assumptions
and, apart from the Lauer \& Postman measurement, they are broadly
consistent with one another, although the Mould \etal measurement is
somewhat high.  The data from the first three surveys are broadly
consistent with the predictions of all our models except \SCDM\ which
produces velocities about factor of 2 lower than the data on large
scales. None of the models is consistent with the measurement of 
Lauer \& Postman who inferred a bulk flow of $764\pm 160\,\kms$ (as
reanalysed by \cite{Colless95}) on a scale of $\sim
80\hmpc$ from a sample of brightest cluster galaxies.  The results in the
figure show that bulk flows are insensitive to the value of $\Omega_0$ when
one focusses attention on models that agree with the observed cluster
abundance. If anything, observed bulk flows constrain the shape of the
power spectrum on large-scales or, in the case of the Lauer \& Postman
result, they conflict with the entire class of models we are considering.

\subsection{Pairwise velocities\label{velco-section}}

We now consider the lower order moments of the pairwise velocity
distribution of dark matter particles in our four cosmological models.
Specifically, we consider the following quantities: \vrad, the mean radial
peculiar velocity of approach between particle pairs; \vradisp, the
dispersion in the radial velocities of pairs; and \vtandisp, the dispersion
in the mean transverse relative velocities of pairs. Following standard
practice, \vradisp\ is not centered; to center one just needs to subtract
\vrad\ in quadrature. These quantities are not directly observable, but we
also compute the dispersion, $\sigma^2_{los}$, the line-of-sight velocity
dispersion (this time centered), defined as:
\begin{equation}
     \sigma^2_{los}(r) = {\int
          \xi(R)\;  \sigma^2_{\rm proj}(R)\; \ddd l \over 
          \int \xi(R)\;\ddd l}
\label{los_equation}
\end{equation}
where $r$ is the projected separation, $R=\sqrt{r^2 + l^2}$, and the
the integral is taken along the line-of-sight between $\pm25\hmpc$.
The quantity $\sigma^2_{\rm proj}$ is the line-of-sight centred 
pairwise dispersion
which is given by:
\begin{equation}
     \sigma^2_{\rm proj} = {r^2v_{\perp}^2/2 + l^2(v_{\parallel}^2 - v^2_{21})
      \over r^2 + l^2}
\end{equation}
This quantity is somewhat
closer to measurements accesible in galaxy redshift surveys; it is a
much weaker function of apparent separation than
\vradisp\ and \vtandisp.

\begin{figure}
\plotone{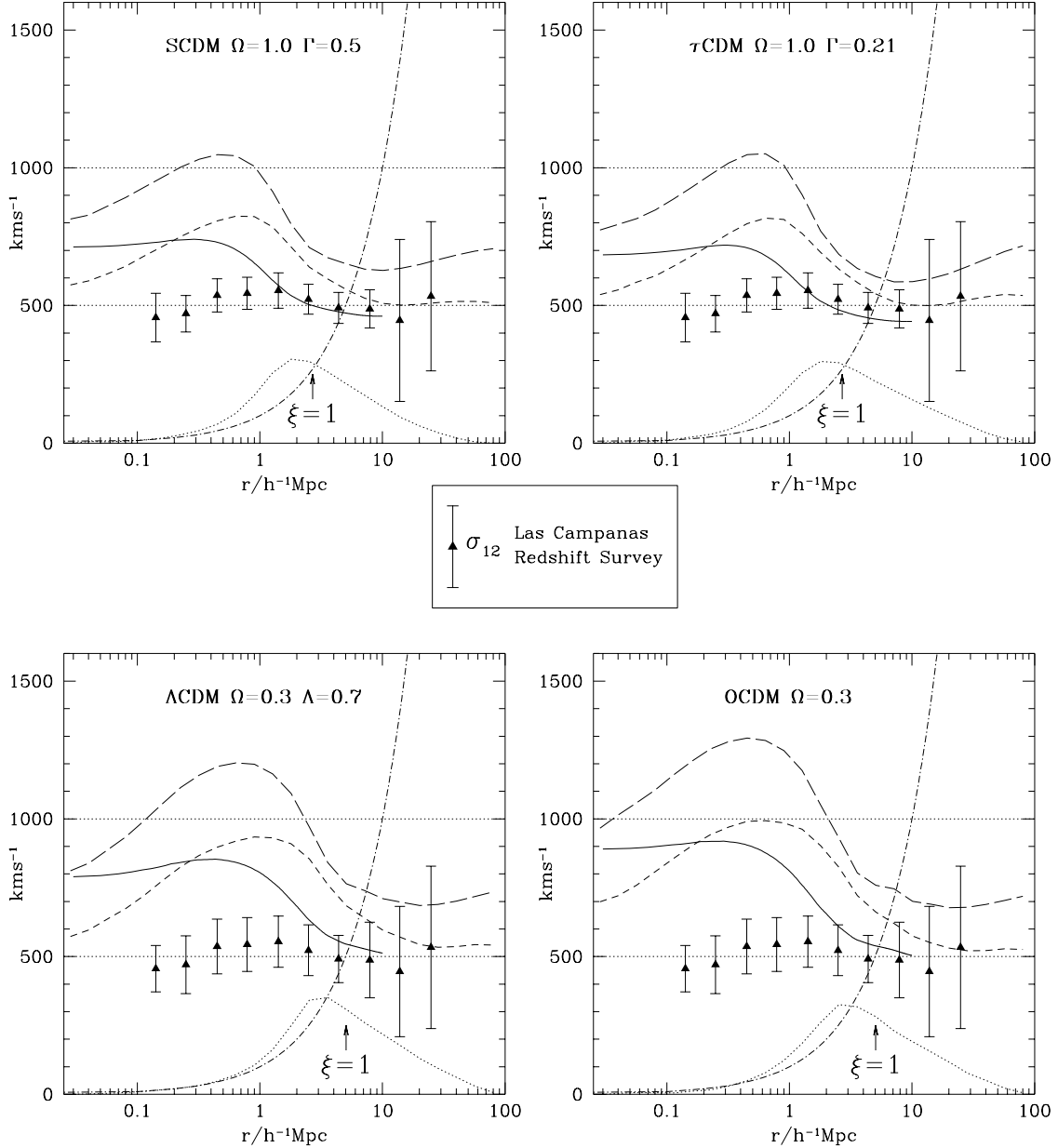}
\caption{Pairwise velocity statistics. In each panel, the dotted curve is
the mean inward radial velocity \vrad; the short dashed line is the
dispersion in the pairwise radial peculiar velocities \vradispo; the
long dashed line is the dispersion in the relative pairwise tangential
peculiar velocities, \vtandispo; the solid line is the line-of-sight
dispersion, $\sigma_{los}$; and the dot-dashed line is the Hubble line
given by $v_{Hubb} = -Hr$, where $H$ is Hubble's constant $r$ is 
physical separation.  The dispersion \vradispo\ is uncentered; to
center, subtract \vrad\ in quadrature. The data points are taken from
Jing, Mo and B\"orner (1997) and show the pairwise velocity
dispersion, $\sigma_{12}$ estimated for the Las Campanas redshift
survey.  These points should be compared to the line-of-sight
dispersions for the models. See main text for discussion of the error
bars used on these points.}
\label{velco_figure}
\end{figure}

Figure \ref{velco_figure} shows \vrad, \vradispo, \vtandispo\ and
$\sigma_{los}$\ as a function of pair separation in our models. Also drawn
on each panel is the Hubble line, given by $v_{\rm Hubb} = -Hr$, where $H$\
is Hubble's constant and $r$ is pair separation in physical units.  Pairs
at fixed physical separation lie on this line. In the stable
clustering regime (\cite{PJEP80}), \vrad\ must follow $v_{\rm Hubb}$. The
distance at which the mass correlation function equals unity, the
correlation length, is marked by an arrow.

The mean pairwise radial velocities, \vrad, vanish at the smallest
separations resolved in our simulations. In the low-$\Omega_0$ models, where
the growth of structure is freezing out at low redshift, \vrad\ follows the
Hubble line up to scales $\sim 300\hkpc$. This indicates that
structures on these scales have almost completely relaxed and the
clustering is stable. In the $\Omega=1$ models there is still a net radial
inflow on these scales although the inflow timescale is longer than the
Hubble time and very much longer than the local dynamical time of pairs at
these separations (where $\xi(r) = 80-200$); the latter is, in turn, much 
shorter than the Hubble time. The pairwise radial velocity in these models
reaches a peak inside the correlation length (marked by the arrow), around
$2-3\hmpc$. This indicates the typical scale of virialising structures at
$z=0$ in the $\Omega=1$ models. At larger radial separations \vrad\
intersects the Hubble line and, at very large separations, it
decays to zero, in accordance with the principle of large-scale isotropy
and homogeneity.

For the same reasons, one expects the ratio \vtandisp/\vradisp\ to tend to
$\sqrt{2} = 1.414$ at large separations. The measured ratios at a
separation of $80\hmpc$ are 1.38, 1.34, 1.36 and 1.37 for \SCDM, \tCDM,
\LCDM, and \OCDM. At scales of a few $\hmpc$, where radial infall is at
its most important, the ratio in the \SCDM\ model is about 1.23 (after
centering). At smaller scales still, the relative motions inside virialised
structures again become closer to isotropy, in agreement with results from
high resolution simulation of dark halos (\cite{Tor96}, \cite{Thomas97}). 
On very small scales, two-body effects contribute to the
isotropization of the orbits.

As was the case with the mean bulk flows and velocity dispersions in random
spheres discussed in subsection~\ref{bulkflow}, the moments of the pairwise
velocity distribution are very similar in the different cosmologies.
As before, this similarity is a direct consequence of our adopted
normalisation. The largest differences occur between the \OCDM\ and \tCDM\
models on small scales - a difference of about $200\,\kms$ in
$\sigma_{los}$. Qualitatively, the trends seen in Fig~\ref{velco_figure}
agree with the analytical calculation of Mo \etal (1996) who find that
pairwise velocities in open models are slightly larger than in $\Lambda$
models and these, in turn, are larger than in $\Omega=1$ models.

It is difficult to compare the predicted dark matter pairwise
velocities with galaxy measurements for a variety of reasons. Firstly,
the velocity dispersion of the dark matter distribution in the
simulations includes a contribution from the internal dispersion of
virialized halos. Secondly, there is some evidence that the velocity
dispersion of dark halos in simulations may be biased low relative to
the dark matter velocity dispersion even after allowing for
contamination from virialized halos (\cite{Carlberg89}), an effect
which Carlberg, Couchman \& Thomas (1990) argue is due to dynamical
friction (see also \cite{Zurek94}). (The velocities of the dark matter
halos in our simulations will be analysed in a future paper by Frenk
\etal 1997.) Finally, biases in the spatial distribution of galaxies
may introduce further biases in the pairwise velocity statistics of
the galaxies relative to the dark matter (Fisher \etal 1994,
\cite{Weinberg95},
\cite{ESD94}.)

Observationally, the velocity dispersion of galaxy pairs is determined by
fitting a model under certain assumptions regarding the two-point
correlation function and the spatial dependence of the infall velocity and
dispersion (\cite{MDP83}.) These assumptions do not necessarily match the
simulation data. More importantly, as Marzke \etal (1995) and Mo \etal
(1996) have argued, pairwise velocity statistics are not robust when
determined from relatively small redshift surveys since these statistics
contain significant contributions from galaxy pairs in rare, massive
clusters. This is not a problem in our simulations which sample a volume of
$13.8\times 10^6(\hmpc)^{3}$, but it is a problem in the present generation
of redshift surveys with the possible exception of the Las Campanas
Redshift Survey (\cite{Shect96}, hereafter LCRS.) Estimates of the pairwise
velocity dispersion in the LCRS, obtained by Jing \etal (1997), are shown
as data points in Figure~\ref{velco_figure}. The LCRS contains quite a
number of rich clusters and appears to give consistent estimates when split
into Northern and Southern subsamples. The error bars plotted in the figure
are the sum in quadrature of the errors obtained directly from the data by
Jing \etal (1997) plus the 1$\sigma$ uncertainties found from applying the
same estimator to mock catalogues constructed from N-body simulations by
these authors. The LCRS velocities are substantially larger than most
previous determinations. The dispersion remains approximately constant over
the range $0.15 -10\hmpc$, reaching an amplitude of $570\pm80\,\kms$ at
$1\hmpc$.

The LCRS data may be compared with the line-of-sight dispersions plotted
for each of our simulations in Figure~\ref{velco_figure}.  At pair
separations $\gtrsim2\hmpc$, all our models are consistent with the data,
although the low-$\Omega$ models lie somewhat low. At smaller separations,
all model curves rise above the data. This difference in behavior may be
due, in part, to the different methods for estimating the dispersion in the
simulations and the data, but it very likely reflects also the biases
present in the simulations mentioned earlier. Interestingly, the
$\Omega=1$ models are closer to the data on small scales than the
low-$\Omega$ models, implying that substantially stronger velocity biases
are required in low-$\Omega$ models to bring them into agreement with the
data.

\section{Discussion and conclusions\label{Summary}}

We have used a suite of high resolution N-body simulations to investigate
the clustering evolution of dark matter in four different cold dark matter
cosmologies. Our simulations followed approximately 17 million
particles. Most of our analysis is based on simulations of very large
cosmological volumes ($239.5\hmpc)^3$, but we also analysed simulations of
somewhat smaller volumes and correspondingly higher mass resolution. The
large volumes and particle numbers, together with a relatively small
gravitational softening ($\sim 30 \hkpc$), allow us to calculate the
clustering and kinematical properties of the dark matter with unprecedented
accuracy. For example, we are able to determine the mass autocorrelation
function over nearly 3 decades in pair separation with better accuracy than
in previous simulations and also with higher precision than is attainable
with existing or planned surveys of galaxies. Our model mass correlation
functions are well fit by an analytic model of the type proposed by
Hamilton \etal (1991) but with the form and parameters proposed by Peacock
\& Dodds (1996). This model may therefore be used to extend some of the
results of our analysis to cosmologies with different parameter values to
those assumed in our simulations.

Two of the four variants of the CDM cosmology that we have investigated are
motivated by various lines of astronomical evidence which suggest a low
cosmological density parameter, $\Omega_0 \simeq 0.3$, and a spectral shape
parameter, $\Gamma=0.21$; we study both a flat model with a non-zero
cosmological constant (\LCDM) and an open model (\OCDM). The remaining two
models both have $\Omega=1$, but one has the standard power spectrum
(\SCDM) and the other has $\Gamma=0.21$ (\tCDM). In all cases, we have
chosen to normalise the primordial fluctuation spectrum so that the present
abundance of rich clusters is approximately reproduced in all the models.
We regard this choice as preferable to the often used alternative of
normalising to the amplitude of the COBE microwave background
anisotropies. With standard assumptions (a Harrison-Zeldovich primordial
spectrum and no contribution to the anisotropy from tensor modes), the
cluster normalisation is close to the COBE normalisation for the \LCDM\ and
\tCDM\ models, but it is significantly higher for the \OCDM\ and
significantly lower for the \SCDM\ model.  With our choice of
normalisation, the overall appearance of all models is determined primarily
by their $\sigma_8$ values with the result that the two high density models
look very similar while the two low density models show more structure but
resemble each other closely.

Our main results concern the detailed properties of the spatial
distribution and velocity fields of the dark matter at $z=0$.  We now
discuss our results and display them concisely in Table~2. 
In all the models the shape of the two-point
correlation function, $\xi(r)$, and power spectrum, $\Delta^2(k)$, of
the dark matter differ significantly from those of the observed galaxy
distribution. In particular, they fail to reproduce the accurate
power-law which the APM survey (and others before that;
c.f. \cite{GP77}) exhibits over nearly four orders of magnitude in
amplitude. At small, but still well-resolved pair separations, all our
model correlation functions become shallower, while at intermediate
separations they all have an inflection point. Uniquely amongst the
models we have explored, \tCDM\ has a mean correlation slope which is
approximately correct over the bulk of the observable range, but even
in this case there are substantial discrepancies on scales smaller
than $\sim 0.2\hmpc$.  Thus, for any of these models to provide an
acceptable representation of reality, the distribution of galaxies
would need to be biased relative to the mass in a non-trivial,
scale-dependent, fashion. Whatever the processes involved in biasing
the galaxy distribution may be, they must conspire to iron out the
features in the dark matter correlation function.

\placetable{tbl-2}

We define a ``bias function'' as the square root of the the ratio of the
galaxy to the mass autocorrelation functions. Our simulations, together
with the galaxy autocorrelation function measured from the APM survey by
Baugh (1996), give the bias as a function of scale accurately for the four
models we have investigated. We find that our two $\Omega=1$ models require
a bias greater than unity everywhere. In the \SCDM\ case, the bias grows
from $\sim 1$ at $\sim 1\hmpc$ to $\sim 1.5$ at $\sim 8\hmpc$ and rises
sharply beyond that. In the \tCDM\ model the bias is approximately
constant, at $b\simeq 1.5$, between $\sim 0.2 \hmpc$ and $\sim 20 \hmpc$.

By design, our low-$\Omega_0$ models have a power spectrum that
approximates that of the APM galaxy survey on large scales. However, even
in this case, the match is not perfect and some amount of bias may still be
required at separations $>10 \hmpc$. Furthermore, these models have the
undesirable feature that the mass correlation function rises above the APM
galaxy correlation function at pair separations smaller than $\sim 5
\hmpc$. On these scales, an ``antibias'' is required for these models to
match the observed galaxy clustering. Galaxy mergers in high density
regions may plausibly suppress small-scale correlations, but it remains to
be seen whether an antibias of the required magnitude is achievable in
practice. Antibiasing may be difficult to reconcile with observed cluster
mass-to-light ratios. In standard virial analyses of clusters, a value of
$\Omega_0$ is derived from the measured mass-to-light ratio by assuming
that the galaxies cluster just like the mass.  With this assumption
Carlberg \etal (1997), for example, inferred $\Omega_0= 0.19\pm0.06$ from
the CNOC sample of intermediate redshift clusters. If galaxies were
actually antibiased, this estimate of $\Omega_0$ would need to be corrected
downwards. However, models with lower values of $\Omega_0$ require higher
values of $\sigma_8$, and even stronger antibias, in order to reproduce the
observed abundance of clusters.

Our simulations allow us to calculate accurately the velocity fields of the
dark matter over a wide range of scales. These are very similar in all our
models, whether they be characterised as bulk flows, single-particle or
pairwise velocity dispersions. This similarity in the velocity fields is a
direct consequence of our adopted normalisation and runs contrary to the common
belief that the amplitude of the observed galaxy velocity fields can be
used to constrain the value of $\Omega_0$. A residual dependence of the
velocity field on the shape of the power spectrum causes the velocities in
the \SCDM\ model to be somewhat lower than in the other models, but amongst
the latter there is no discernible difference. For example, the 1D velocity
dispersion of the dark matter is approximately $600\,\kms$ in all the
models, and the line-of-sight pairwise velocity dispersions fall in the
range $700-900\,\kms$. The first of these numbers is reminiscent of the
peculiar velocity of the Local Group, while the second is consistent with,
although on the high side of, a recent determination from the Las Campanas
redshift survey at a pair separation of $\sim 1\hmpc$ (Jing \etal 1997). On
smaller scales, our simulations, particularly our low-$\Omega_0$ models,
predict higher pairwise velocity dispersions than inferred from this
survey, indicating that a substantial velocity bias is required  
to bring the models into agreement with the data. Bulk flows on
large-scales are most accurately calculated using linear theory. Our models
all predict somewhat smaller values than those estimated from recent
surveys of the local universe (\cite{Mould93}; \cite{Courteau93}; Dekel
\etal\ 1997) but, with the exception of \SCDM, they are consistent with
these data. None of the models reproduces the large bulk flows inferred by
Lauer \& Postman (1994).

High resolution simulations like those presented here allow very accurate
measurements of the clustering distribution of dark matter. Further
progress in this subject will rely on the ability to address the outstanding
issue that limits the comparison of these models with observations: the
connection between the distribution of mass and the distribution of
galaxies. This will require a realistic treatment of the evolution of the
baryonic component of the Universe.

\acknowledgments 

We are grateful to Carlton Baugh for useful discussions and for providing
us with the APM galaxy survey data used in Figures~\ref{correl_fig}
and~\ref{power_figure}. We thank David Weinberg for suggesting several
significant improvements to the manuscript and Avishai Dekel for
communicating to us, in advance of publication, results of his bulk
analysis shown in Figure~\ref{bulkfig}. CSF acknowldeges a PPARC Senior
Fellowship. This work was supported in part by grants from PPARC, EPSRC and
the EC TMR network for ``Galaxy formation and evolution.''  The simulations
reported here were carried on the Cray-T3Ds at the Edinburgh Parallel 
Computing Centre and the Rechenzentrum, Garching. We thank the Editor 
for suggesting the inclusion of Table~2.

\clearpage
\appendix
\section{Appendix: Derivation of equation (\ref{eqncorrxi})}
The two-point correlation function is related to the power spectrum
by:
\begin{equation}
   \xi({\bf r}) = \int P({\bf k}) \exp[i{\bf k\cdot r}]\ddd^3{\bf k}, 
\label{eqa1}\end{equation}
where bold font implies that the quantity is a 3-dimensional vector.

In deriving a correction to the linear correlation function for a
periodic box we must make an assumption for how the power selected
for each discrete mode of the periodic box is related to the power
density of the same mode in the continuous power spectrum. As discussed
in Section~\ref{rundetails}, we draw the power for each mode from an
exponential distribution with the mean power set by the power density
of the mode in the continuous power spectrum. Thus, the
ensemble-average linear correlation function of the periodic boxes, 
$\xi_s({\bf r})$, is given by:
\begin{equation}
 \xi_s({\bf r}) =   ({2\pi\over L})^3 \sum_{{\bf b} = (0,0,0)}^\infty
    P({2\pi{\bf b}\over L}) \exp[2\pi i{\bf b\cdot r}/L], 
\label{eqa2}\end{equation}
where $L$ is the simulation boxsize and the sum over ${\bf b}$ is a
sum over all integer triples. The correction we derive is a systematic
correction that applies to an ensemble of simulations. 

We make use of the Poisson summation formula which, for a function 
$\phi({\bf x})$, states that:
\begin{equation}
  \sum_{{\bf b}=(0,0,0)}^\infty \phi\big(2\pi{\bf b}\big)= 
{1\over(2\pi)^3}\sum_{{\bf n}=(0,0,0)}^\infty
   \int \phi({\bf t})\exp[i{\bf n\cdot t}] \ddd^3{\bf t}, 
\end{equation}
subject to certain conditions on the function $\phi({\bf x})$ which hold for
 the case of interest here (see Courant and Hilbert 1953, p.76).

Substituting the r.h.s. of equation~(\ref{eqa2}) into the Poisson summation
formula we obtain: 
\begin{equation}
   \xi_s({\bf r}) = \sum_{{\bf n}=(0,0,0)}^\infty\int P({\bf k}) 
          \exp[i{\bf k\cdot}({\bf r}-L{\bf n})]\ddd^3{\bf k}. 
\end{equation}
From equation~(\ref{eqa1}) we can rewrite this as:
\begin{equation}
 \xi_s({\bf r}) = \xi({\bf r}) + \sum_{{\bf n}\neq(0,0,0)}^\infty 
        \xi({\bf r}-L{\bf n})
\end{equation}
Applying this to the evolved linear power spectrum, which is isotropic, we
arrive at the correction term, eqn~(\ref{eqncorrxi}), to the
correlation function for the periodic box:
\begin{equation}
\bigtriangleup\xi({\bf r}) = \sum_{{\bf n}\neq(0,0,0)}^{\bf\infty}
-\xi_{lin}(|{\bf r} + L{\bf n}|)
\end{equation}

\clearpage
 
\begin{deluxetable}{crrrrrrrrrrr}
\footnotesize
\tablecaption{Cosmological and Numerical Parameters of Runs \label{tbl-1}}
\tablewidth{0pt}
\tablehead{
\colhead{Run} & \colhead{$\Omega_0$}   & \colhead{$\Lambda$} & \colhead{$h$} & 
\colhead{$\Gamma$}  & \colhead{$\sigma_8$} & \colhead{$L/\hmpc$} & 
\colhead{Npar}     & \colhead{$m_p/\hmsun$}  & 
\colhead{$l_{\rm soft}/h^{-1}{\rm Kpc}$}
}
\startdata
\SCDMa  &1.0 &0.0 &0.5 &0.50 &0.51 &239.5\whsp &$256^3$ &$2.27\times10^{11}$ &36\whspb \nl
\TCDMa  &1.0 &0.0 &0.5 &0.21 &0.51 &239.5\whsp &$256^3$ &$2.27\times10^{11}$ &36\whspb \nl
\TCDMab  &1.0 &0.0 &0.5 &0.21 &0.51 &239.5\whsp &$256^3$ &$2.27\times10^{11}$ &36\whspb \nl
\LCDMa  &0.3 &0.7 &0.7 &0.21 &0.90 &239.5\whsp &$256^3$ &$6.86\times10^{10}$ &25\whspb \nl
\OCDMa  &0.3 &0.0 &0.7 &0.21 &0.85 &239.5\whsp &$256^3$ &$6.86\times10^{10}$ &30\whspb \nl
\SCDMb  &1.0 &0.0 &0.5 &0.50 &0.51 &84.5\whsp   &$256^3$ &$1.00\times10^{10}$ &36\whspb \nl
\TCDMb  &1.0 &0.0 &0.5 &0.21 &0.51 &84.5\whsp   &$256^3$ &$1.00\times10^{10}$ &36\whspb \nl 
\LCDMb  &0.3 &0.7 &0.7 &0.21 &0.90 &141.3\whsp   &$256^3$ &$1.40\times10^{10}$ &30\whspb \nl
\OCDMb  &0.3 &0.0 &0.7 &0.21 &0.85 &141.3\whsp   &$256^3$ &$1.40\times10^{10}$ &30\whspb \nl 
\enddata

% Text for table footnotes follows the tabular data and must be inside the
% deluxetable environment.  Note that it is OK to put \ref's in 
% \tablenotetext's.

\end{deluxetable}

\clearpage
\begin{deluxetable}{ccccccc}
\footnotesize
\tablecaption{Summary of Results\label{tbl-2}}
\tablewidth{0pt}
\tablehead{
\colhead{Model\tablenotemark{a}} & \colhead{Cluster} & \colhead{COBE} & \colhead
{Constant}
& \colhead{Small scale} & \colhead{$V_{\rm bulk}$\tablenotemark{c}}
& \colhead{Pairwise} \\
\colhead{} & \colhead{Abundance} & \colhead{Norm} & \colhead{Bias} & 
\colhead{Bias/Anti} & 
\colhead{} & \colhead{Velocities} 
} 
\startdata
\SCDM  &Yes &No   &No  &Bias     &Low                  &Slightly high \nl
\tCDM  &Yes &Yes  &Yes &Bias     &OK                   &Slightly high \nl
\LCDM  &Yes &Yes  &No  &Antibias &OK                   &high          \nl
\OCDM  &Yes &No\tablenotemark{b} &No  &Antibias &OK  &high          \nl
\enddata

% Text for table footnotes follows the tabular data and must be inside the
% deluxetable environment.  Note that it is OK to put \ref's in 
% \tablenotetext's.

\tablenotetext{a}{See table~1 for the definitions of the models.}
\tablenotetext{b}{A model with a $\Omega_0 = 0.4$ and a slightly lower
value of $h$ can agree with both the cluster abundance and COBE DMR
constraints.}
\tablenotetext{c}{When compared to the Dekel \etal\ 1997 data points. All 
the models are strongly inconsistent with the Lauer \&
Postman 1994 result.}
\end{deluxetable}

\end{document}